\documentclass[a4paper,conference]{IEEEtran}
\usepackage[left=1.57cm,right=1.57cm,top=1.52cm,bottom=3cm]{geometry}
\IEEEoverridecommandlockouts

\usepackage{cite}
\usepackage{tikz}
\usetikzlibrary{arrows.meta}
\usepackage{pgfplots}
\usepackage[ruled,vlined]{algorithm2e}
\usepackage{psfrag}
\usepackage{soul}
\usepackage{graphicx}
\usepackage{textcomp}
\usepackage{bbm}
\usepackage{caption}
\usepackage{subcaption}
\usepackage{amsmath,amssymb,amsfonts, acronym, amsthm}
\usepackage{hyperref}
\usepackage{xparse}
\usepackage{float}
\usepackage{amsmath, amssymb, bm, mathrsfs, yfonts}

\acrodef{5G}{fifth generation}
\acrodef{6G}{sixth generation}
\acrodef{AOA}{angle-of-arrival}
\acrodef{AOD}{angle-of-departure}
\acrodef{AI}{artificial intelligence}
\acrodef{UAV-I}{{\em UAV intelligence}}
\acrodef{AI/ML}{artificial intelligence/machine learning}
\acrodef{A2A}{agent-to-agent}
\acrodef{AWGN}{additive white Gaussian noise}
\acrodef{AO}{Alternate Optimization}
\acrodef{AP}{Access Point}
\acrodef{BS}{base station}
\acrodef{BCD}{Block Coordinate Descent}
\acrodef{CRLB}{Cram\'er-Rao Lower Bound}
\acrodef{CDF}{cumulative density function}
\acrodef{CDR}{correct detection rate}
\acrodef{CIR}{channel impulse response}
\acrodef{CRAS}{connected robotics and autonomous systems}
\acrodef{CNPC}{control and non-payload communication}
\acrodef{CSI}{Channel State Information}
\acrodef{DMA}{dynamic metasurface antenna}
\acrodef{CR}{Constraint Relaxation}
\acrodef{DEC-POMDP}{decentralized partially observable Markov decision process}
\acrodef{D2D}{device-to-device}
\acrodef{DC}{dual connectivity}
\acrodef{DL}{downlink}
\acrodef{EM}{electromagnetic}
\acrodef{EKF}{extended Kalman filter}
\acrodef{EIRP}{effective isotropic radiated power}
\acrodef{FAR}{false alarm rate}
\acrodef{FIM}{Fisher Information matrix}
\acrodef{FL}{federated learning}
\acrodef{FMCW}{frequency modulated continuous wave}
\acrodef{FDMA}{Frequency Division Multiple Access}
\acrodef{FD}{Full Duplex}
\acrodef{FSS}{frequency selective surface}
\acrodef{GLRT}{generalized likelihood ratio test}
\acrodef{GP}{Gaussian process}
\acrodef{GPI}{generalized policy iteration}
\acrodef{GAM}{gradient ascent method}
\acrodef{IoT}{Internet-of-Things}
\acrodef{IS}{image similarity}
\acrodef{IRS}{intelligent reflecting surfaces}
\acrodef{IO}{Indoor Office}
\acrodef{KPI}{key performance indicator}
\acrodef{KKT}{Karush Kuhn Tucker}
\acrodef{KF}{Kalman filter}
\acrodef{LLRT}{log-likelihood ratio test}
\acrodef{LOS}{line-of-sight}
\acrodef{LS}{least square}
 \acrodef{MAC}{medium access control}
\acrodef{MAP}{maximum a-posteriori probability}
\acrodef{MEC}{multi-access edge computing}
\acrodef{MAB}{multi-armed bandit}
\acrodef{mMTC}{massive machine type communication}
\acrodef{eMBB}{enhanced mobile broadband}
\acrodef{MMSE}{minimum mean squared error }
\acrodef{MARL}{multi-agent reinforcement learning}
\acrodef{MDP}{Markov decision process}
\acrodef{MIMO}{multiple input multiple output}
\acrodef{ML}{machine learning}
\acrodef{MLE}{maximum likelihood estimator}
\acrodef{mm-wave}{millimeter-wave}
\acrodef{MISO}{Multiple Input Single Output}
\acrodef{MSE}{Mean Square Error}
\acrodef{MRT}{Maximum Ratio Transmission}
\acrodef{MM}{Minorization-Maximization}
\acrodef{MTP}{metaprism}
\acrodef{NOMA}{non-orthogonal multiple access}
\acrodef{NLOS}{non line-of-sight}
\acrodef{OG}{occupancy grid}
\acrodef{OFDM}{orthogonal frequency division multiplexing}
\acrodef{OFDMA}{orthogonal frequency division multiple access}
\acrodef{PF}{particle filtering}
\acrodef{pdf}{probability density function}
\acrodef{PFA}{probability of false alarm}
\acrodef{PL}{packet loss}
\acrodef{POMDP}{partially observable Markov decision process}
\acrodef{PER}{packet error rate}
\acrodef{PEB}{position error bound}
\acrodef{PD}{Path Discretization}
\acrodef{QoS}{quality of service}
\acrodef{RIS}{reconfigurable intelligent surface}
\acrodef{RCS}{radar cross section}
\acrodef{RMSE}{root mean square error}
\acrodef{RFID}{radiofrequency identification}
\acrodef{RL}{reinforcement learning}
\acrodef{ROC}{receiver operating characteristics}
\acrodef{RR}{reading range}
\acrodef{RRCS}{root-radar cross section}
\acrodef{RV}{random variable}
\acrodef{SLAM}{simultaneous localization and mapping}
\acrodef{SNR}{signal-to-noise ratio}
\acrodef{SIR}{sequential importance resampling} 
\acrodef{SOCP}{second-order cone programming}
\acrodef{SDR}{Semidefinite Relaxation}
\acrodef{SM}{Shopping Mall}
\acrodef{SINR}{signal-to-interference-plus-noise ratio}
\acrodef{SE}{Spectral Efficiency}
\acrodef{SCA}{Successive Convex Approximation}
\acrodef{STAR}{simultaneously transmitting and reflecting}
\acrodef{SIC}{Successive Interference Cancelation}
\acrodef{SDP}{Semidefinite Programming}
\acrodef{SISO}{single input single output}
\acrodef{TD}{temporal-difference}
\acrodef{TOA}{time-of-arrival}
\acrodef{TDMA}{Time Division Multiple Access}
\acrodef{UE}{User Equipment}
\acrodef{THz}{Terahertz}
\acrodef{UAV}{unmanned aerial vehicle}
\acrodef{U2U}{UAV-to-UAV}
\acrodef{UKF}{Unscented Kalman Filter}
\acrodef{UCB}{upper confidence bound}
\acrodef{URLLC}{ultrareliable and low-latency communication}
\acrodef{UL}{uplink}
\acrodef{VLC}{visible light communication}

\acrodef{WMMSE}{weighted Minimum Mean Square Error}

\usepackage{color}
\usepackage[markup=underlined,ulem=normalem,xcolor=red]{changes}
\usepackage[singlespacing]{setspace} 

\definecolor{silver}{rgb}{0.95, 0.95, 0.95}

\IEEEoverridecommandlockouts

\usetikzlibrary{positioning} 
\definecolor{Eored}{rgb}{.647 ,.129 ,.149} 
\definecolor{Eogreen}{rgb}{0 ,0.53 ,0}

\definechangesauthor[name=<name>, color=orange]{AT}

\newcommand{\Gn}{\Gamma_{n}}
\newcommand{\Xn}{X_{n}}
\newcommand{\Pn}{\psi_{n}}

\newcommand{\tinc}{\theta_{\text{inc}}}

\newcommand{\tk}{\theta^{(k)}}

\newcommand{\bXn}{\bar{X}_{n}}
\newcommand{\bXnp}{\bar{X}_{n,p}}
\newcommand{\bLnp}{\bar{L}_{n,p}}
\newcommand{\bCnp}{\bar{C}_{n,p}}
\newcommand{\betapn}{\beta_{n,p}}

\newcommand{\an}{\alpha_{n}}
\newcommand{\gn}{\gamma_{n}}

\newcommand{\M}{\mathcal{M}}

\newcommand{\tm}{\theta_m}
\newcommand{\tM}{\theta_M}

\newcommand{\Poln}{P_{n}}
\newcommand{\fk}{f_k}

\newtheorem{lemma}{Lemma}
\newtheorem{definition}{Definition}
\begin{document}
	\author{Silvia~Palmucci \IEEEmembership{Student~Member,~IEEE}, Andrea~Abrardo \IEEEmembership{Senior~Member,~IEEE},\\ Davide Dardari \IEEEmembership{Fellow,~IEEE}, Alberto Toccafondi \IEEEmembership{Senior~Member,~IEEE},\\ Marco~Di~Renzo \IEEEmembership{Fellow,~IEEE}
		\thanks{S. Palmucci, A. Abrardo, A. Toccafondi are with the University of Siena and CNIT, Italy (e-mail: silvia.palmucci@student.unisi.it, abrardo@dii.unisi.it, alberto.toccafondi@unisi.it). D. Dardari is with the WiLAB - Department of Electrical and Information Engineering ``Guglielmo Marconi" - CNIT, University of Bologna, Italy (e-mail: davide.dardari@unibo.it). M. Di Renzo is with Universit\'e Paris-Saclay, CNRS, CentraleSup\'elec, Laboratoire des Signaux et Syst\`emes, 3 Rue Joliot-Curie, 91192 Gif-sur-Yvette, France. (marco.di-renzo@universite-paris-saclay.fr).
			The work of D. Dardari was supported in part by the European Commission through the Project HORIZON-JU-SNS-2022 project TIMES (Grant no. 101096307) and under the Italian National Recovery and Resilience Plan (NRRP) of NextGenerationEU, partnership on ``Telecommunications of the Future” (PE00000001 - program ``RESTART”).
			The work of M. Di Renzo was supported in part by the European Commission through the Horizon Europe project titled COVER under grant agreement number 101086228, the Horizon Europe project titled UNITE under grant agreement number 101129618, and the Horizon Europe project titled INSTINCT under grant agreement number 101139161, as well as by the Agence Nationale de la Recherche (ANR) through the France 2030 project titled ANR-PEPR Networks of the Future under grant agreement NF-PERSEUS 22-PEFT-004, and by the CHIST-ERA project titled PASSIONATE under grant agreements CHIST-ERA-22-WAI-04 and ANR-23-CHR4-0003-01.
		}
	}
	\title{Metaprism Design for Wireless Communications: Angle-Frequency Analysis, Physical Realizability Constraints, and Performance Optimization}
	
	\maketitle
	\begin{abstract}
		Recent advancements in smart radio environment technologies aim to enhance wireless network performance through the use of low-cost electromagnetic (EM) devices. Among these, reconfigurable intelligent surfaces (RIS) have garnered attention for their ability to modify incident waves via programmable scattering elements.
		{An RIS is a nearly-passive device, in which the tradeoff between performance, power consumption, and optimization overhead depend on how often the RIS needs to be reconfigured.} 
		This paper focuses on the \ac{MTP}, a {static} frequency-selective metasurface {which relaxes the reconfiguration requirements of RISs and allows for the creation of different beams at various frequencies. }In particular, we address the design of an ideal \ac{MTP} based on its frequency-dependent reflection coefficients, defining the general properties necessary to achieve the desired beam steering function in the angle-frequency domain. We also discuss the limitations of previous studies that employed oversimplified models, which may compromise performance. Key contributions include a detailed exploration of {the equivalence of the MTP }to an ideal S-parameter multiport model and an analysis of its implementation using Foster{'s} circuits. Additionally, we introduce a realistic multiport network model that incorporates aspects overlooked by {ideal scattering} models, along with an ad-hoc optimization strategy for this model. The {performance} of the proposed optimization approach and circuits implementation are validated through simulations using a commercial full-wave EM simulator, confirming the effectiveness of {the proposed} method.
	\end{abstract}
	\acresetall
	
	\section{Introduction}\label{Intro}
	
	In recent years, extensive research has focused on developing technologies that enable the smart radio environment paradigm, where the propagation environment is adapted to enhance wireless network performance through the deployment of low-cost \ac{EM} devices. Among these, \acp{RIS} have garnered significant attention. These versatile \ac{EM} devices modify incident waves using programmable scattering elements, employing either conventional antenna elements or advanced metasurfaces. {By dynamically adjusting their reflection or transmission characteristics, \acp{RIS} provide a more efficient solution with lower complexity and energy consumption compared to the deployment of additional base stations (BS) or active relays \cite{RenzoZDAYRT20}}.
	
	RISs can operate by processing {impinging waves in space, time and frequency. }The concept of \ac{FSS} has been {studied} for several decades, and many works have contributed to the development {of this technology} \cite{munk2005frequency}. 
	{In early works, the main use of an \ac{FSS} was, in general, to perform spectral filtering in microwave and optical frequency bands through the design of extremely dispersive reflection or transmission properties. Some practical examples include dual-band reflectarrays, antenna covering surfaces, and frequency-selective absorbers.}
	More recently, the frequency selectivity of a metasurface has been exploited as a practical means to change the phase shift applied by each element composing the metasurface.
	An example is given by \acp{DMA}{, which can be implemented through a surface with} sub-wavelength elements operating as resonant circuits with programmable coefficients \cite{shlezinger2021dynamic}. 
	Some representative examples of RIS and DMA implementations that exploit the frequency-selectivity are reported in \cite{wang2020dynamic,Zha:22, yang2020intelligent,zheng2019intelligent, katsanos2022wideband,cai2022irs}.
	{In \cite{dardari2021using}, a frequency-selective metasurface is proposed, which is capable of generating different beams at various frequencies. This technology, which is referred to as \ac{MTP}, provides coverage over wide angular areas by reflecting different sub-carriers of an impinging signal—such as a group of subcarriers in an \ac{OFDMA} signal—toward distinct directions.}
	The \ac{MTP}{, as presented in \cite{dardari2021using}, is a static metasurface that is not reconfigurable and is hence passive and controls} the {scattering} of {impinging waves} without requiring control channels or dedicated \ac{CSI}. {User management is handled by the base station {by controlling the direction and the frequency of the incident signals, and hence the users being served.} {In RIS-aided networks, user assignment can be performed by reconfiguring the metasurface, but \ac{CSI} and a control channel are required, which allow an RIS to be optimized at different time scales} \cite{panreconfigurable2021, zhitwotimescale2023}.}
	{Recent implementations of RISs may relax the need of a centralized control channel \cite{albanese2022Marisa}, but at least long-term \ac{CSI} is required \cite{AbrardoS,zhitwotimescale2023}.}
	
	{In summary, an MTP is a smart skin \cite{oliveri2021holographic} that exploits the frequency selectivity of a metasurface to direct impinging signals towards desired directions of reflection. MTPs} are specifically tailored for scenarios in which a large number of devices must be served simultaneously, each requiring  low data rate and low latency, as required by, e.g., industrial \ac{IoT}. {In fact, the many sub-carriers of an incident \ac{OFDM} signal may be directed towards specified directions of reflection. In general, it is, however, not possible to direct several sub-carriers towards the same direction because of the inherent frequency selective response of a metasurface.} 
	An example of application of this operating principle is investigated in \cite{lotti2023metaprism}, where an \ac{MTP} is exploited to detect the position of a target hidden by an obstacle, using a standard frequency modulated continuous radar signal. 
	A similar approach for localization in \ac{NLOS} conditions using \ac{OFDM} signals is presented in \cite{LotCalDar:C24}. 
	
	The design criterion for the \ac{MTP} proposed in \cite{dardari2021using,lotti2023metaprism,LotCalDar:C24} capitalizes on the {frequency selectivity of the reflection coefficient of appropriately designed unit cells}. 
	{More precisely, the reflection coefficient depends on the surface impedance {of the equivalent circuit }that models the physical implementation of each unit cell \cite{Movahediqomi2023Comparison}. The surface impedance is, in general, frequency dependent, making the reflection coefficient, in amplitude and phase, frequency dependent. } 
	{In \cite{dardari2021using}, the response of each unit cell is modeled as that obtained by an equivalent model consisting of a scatterer (elementary radiator) with an accessible port loaded by a cell-dependent and frequency-dependent impedance. An LC circuit is chosen as an example of frequency-dependent load.
		The LC circuit allows for the realization of the desired impedance in a very narrow range around a central frequency, where the reflection coefficient of the circuit can be linearized. This means that the model with LC circuits is applicable only for very limited bandwidths, i.e., for a small number of sub-carriers around the reference sub-carrier.}
	Furthermore, the {equivalent scattering} model utilized in \cite{dardari2021using, lotti2023metaprism, LotCalDar:C24} does not account for two aspects that play an important
	role in characterizing the operation of a realistic reflecting metasurface: the specular component associated with the structural scattering of the MTP and the \ac{EM} mutual coupling between closely spaced unit cells \cite{RenzoZDAYRT20,Movahediqomi2023Comparison}. {Upon considering the unit cells' equivalent {scattering} model, }such aspects can be adequately taken into consideration adopting multiport network models for the MTP. The initial multiport model for RIS-aided channels, which is non-linear with respect to the reflection coefficient, was introduced in \cite{DR1}, {where the authors {modeled the unit cells {response} of a metasurface by using thin wire dipoles, motivated by the discrete dipole approximation.}
	}A similar approach, based on the coupled dipoles formalism, is discussed in \cite{FaqiriSASIH23}. The multiport model has since been applied to optimize \ac{SISO} and \ac{MIMO} systems, and different channel models \cite{DR2,ABR_MUTUAL}. 
	{In this paper, we utilize the multiport network model for the synthesis of load impedances of equivalent circuits to meet the requirements of an MTP working on a generic band and with the capability to span a generic range of angles.} 
	
	\section{Contribution}
	
	Given the {low complexity, overhead, power consumption, and several potential applications of MPT for communication and sensing}, this work delves into practical models and designs of \acp{MTP} by introducing a more accurate model and suitable optimization methods {for the equivalent MTP {scattering} model.} 
	Specifically, the innovative contributions of this paper are the following:
	\begin{itemize}
		\item The concept of MTP is developed and a design for {an ideal frequency selective }model is proposed. Additionally, a general definition and an angle-frequency analysis are provided to calculate resolution in both domains and to assess the performance in the presence of multipath. 
		\item {Considering an ideal model and assuming the use of dipoles as radiation elements, we study the design of the unit cells to implement the \ac{MTP}. }It is demonstrated that, regardless of the angle-frequency mapping adopted, the realization of the \ac{MTP} requires that the circuits at its ports have poles appropriately positioned in the band of interest, with their number and position varying according to the index of the \ac{MTP} element and the required angular spreading.
		\item {A circuital realization of the {frequency and cell-dependent impedance} of each unit cell }is proposed, which, thanks to an optimization procedure tailored for this purpose, approximates the ideal response with sufficient accuracy.
		\item Building on recent findings (e.g., \cite{abrardo2023design} for \acp{RIS}), we propose a practical {multi-port S-parameter} network model for the MTP that accounts for real-world effects overlooked by the ideal model. We then present a numerical optimization framework that allows implementations using Foster circuits. 
		\item The capabilities of the proposed design to realize the \ac{MTP} are assessed through the use of a commercial full-wave \ac{EM} simulator based on the method of moments.
	\end{itemize}

	\subsection{Paper Outline and Notations}
	The rest of the paper is organized as follows. Sections \ref{sec:SystemModel}-\ref{IDEAL_MTP} introduce the considered system model and the \ac{MTP}, and in Section \ref{sec:CircuitModelDesign} the circuit model design to implement the \ac{MTP} functionalities is analyzed. In Section \ref{sec:RealisticSurfaceModels} a more realistic \ac{MTP} model is considered and the corresponding design is optimized. Sections \ref{sec:Results}-\ref{sec:Conclusions} present numerical results and draw final conclusions.
	As for the notation, matrices, vectors and scalars are denoted by bold uppercase letters, bold lowercase letters and normal fonts, respectively, i.e., $\mathbf{X}$, $\mathbf{x}$, and $x/X$. The notation $\boldsymbol{X} = \{x\}$ is used to indicate that the matrix $\boldsymbol{X}$ is made by all the elements $x$.
	$(\cdot)^{-1}$ and $(\cdot)^{T}$ stand for the inverse and the transpose of the argument, respectively, $\mod(a,b)$ refers to the modulus operation between $a$ and $b$, $\lceil\cdot \rceil$ and $\lfloor\cdot \rfloor$ refer to the ceiling and floor operations of the argument. $\mathcal{CN}(\mu, \sigma^2)$ denotes a complex Gaussian distribution with mean $\mu$ and variance $\sigma^2$, while $\mathbb{E}[\cdot]$ evaluates the expected value of the argument.

	\section{System Model}
	\label{sec:SystemModel}
	
	\begin{figure}
		\vspace{-\baselineskip}
		\centering
		\includegraphics[width =0.6\linewidth]{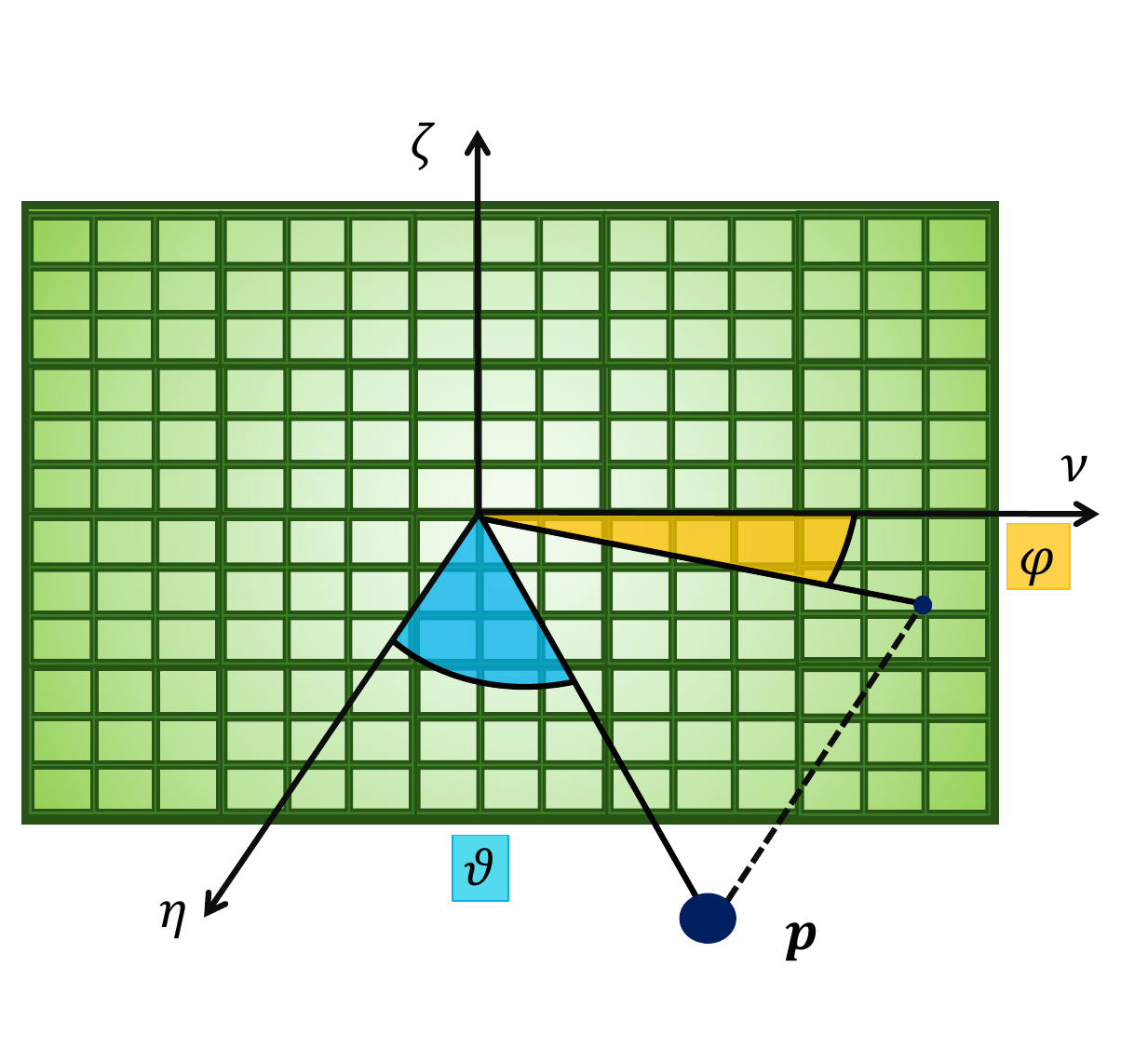}
		\caption{3D coordinate system for an \ac{MTP} placed in the $(\nu, \zeta)$ with normal along $\eta$.}
		\label{fig:assi}
		\vspace{-\baselineskip}
	\end{figure}
	We consider the downlink of a communication scenario where a \ac{BS} located in $\textbf{p}_{BS}$ communicates with one or more nodes, supported by a {frequency-selective metasurface,  referred here as the \ac{MTP}}. 
	{As in \cite{dardari2021using}, the \ac{MTP} is composed of $N = I \times J$ cells. It }is centered in $\textbf{p}_M$ and 
	arranged along two directions $(\nu, \zeta)$, with ${\nu, \zeta} \in \{x, y, z\}$. Each {cell}, of size $d_{\nu} \times d_{\zeta}$, is arranged along spacings of $\Delta_{\nu}$ and $\Delta_{\zeta}$ in the $\nu$ and $\zeta$ directions, respectively. {For $n = 1,\ldots,N$, the location of the $n$-th {cell} of the \ac{MTP} can be expressed as $\mathbf{p}_{n} = \mathbf{p}_{0} + \left(\nu_n \Delta{\nu}, \zeta_n \Delta_{\zeta} \right)$, where $\nu_n = \mod(n-1,I)$, $\zeta_n = \lceil n/I\rceil -1$ and  $\mathbf{p}_0$ being the point on the \ac{MTP} with the smallest $\nu$ and $\zeta$ coordinates, placed at the lower left corner.} It is assumed no direct link between the BS and the node, with communication only occurring through the BS-MTP and MTP-node links. Hence, the BS and the nodes are assumed to be in \ac{LOS} condition with respect to the \ac{MTP}. 
	\subsection{S-parameters Channel Model}
	{{In the following} we consider the equivalent {scattering model} implementation of the MTP's unit cells.} {According to this model each scattering element includes a radiation element, an accessible port, and a frequency dependent static load. The whole analysis is carried out }
	leveraging the S-parameter model discussed in \cite{DR1}. 
	Assuming, for simplicity, a \ac{SISO} scenario, the system transfer function from the incident wave at the transmitter port to the outgoing wave at the receiver port is expressed as 
	\begin{align}\label{eq_model_2}
		&{h}(f) = s_{RT}{(f)} + \mathbf{s}_{RM}{(f)}\left[\boldsymbol\Gamma(f)^{-1}-\mathbf{S}_{SS}{(f)}\right]^{-1}\mathbf{s}_{MT}{(f)},
	\end{align} 
	{where $f$ denotes the frequency. The scalar term $s_{RT}{(f)}$ in \eqref{eq_model_2} represents the system transfer function when all the MTP's ports are matched-terminated. In the absence of a direct link, this term solely depends on the structural scattering of the MTP, making it independent of {the loads at the MTP ports.}} 
	{The terms $\mathbf{s}_{MT}{(f)} \in \mathbb{C}^{N \times 1}$ and $\mathbf{s}_{RM}{(f)} \in \mathbb{C}^{1 \times N}$ are the scattering sub-vectors that describe the relationship between: (i) the incident wave at the transmitter port to the outgoing waves at the MTP ports and (ii) the incident waves at the MTP ports  to the outgoing wave at the receiver port, respectively.}
	The matrix $\boldsymbol{\Gamma}(f) \in \mathbb{C}^{N \times N}$ is a diagonal matrix with $\left\{\boldsymbol{\Gamma}\right\}_{n,n}(f) = \Gamma_{n}(f)$, where $\Gamma_{n}(f)$ is the reflection coefficient, defined as follows: 
	\begin{equation}
		\label{reflection_coeff}
		\Gamma_{n}(f) = e^{\jmath \Pn(f)} = \frac{j\Xn(f) - Z_0}{j\Xn(f) + Z_0}.
	\end{equation}
	$\Xn(f)$ represents the {reactive} load {impedance} of the $n$-th MTP port and $Z_0$ is the reference impedance, typically assumed equal to $50$ $\Omega$. Finally, the matrix $\mathbf{S}_{SS}{(f)}$ is the scattering matrix of the \ac{MTP}, the entries of which are the reflection coefficients of the \ac{MTP} matched to the characteristic impedance $Z_0$ on the main diagonal, and the mutual coupling between the various elements of the \ac{MTP} on the off-diagonal elements.
	Note that in the S-parameter model in \eqref{eq_model_2} 
	the parameters depend on the position of the elements of the \ac{MTP}, the transmitter, and the receiver, and on the type of elementary scatterer considered, such as dipoles or patches. This model is thus a general model that remains valid under both near-field and far-field conditions.

	\subsection{Far-field Channel for an Ideal MTP}
	
	In this section, we introduce a channel model suitable for a far-field scenario with an ideal MTP. This model is widely used in the literature for the characterization of intelligent surfaces and can be seen as a particular case of the model in \eqref{eq_model_2}.\\
	{To begin, we consider an ideal MTP where $\mathbf{S}_{SS}{(f)} = \mathbf{0}$, corresponding to the case where the input impedance at all MTP ports is equal to $Z_0$ and there is no mutual coupling between the MTP elements. Furthermore, we consider the case $s_{RT}{(f)}=0$, which corresponds to the absence of direct link between transmitter and receiver, as well as the absence of structural scattering.} In this case, \eqref{eq_model_2} simplifies to:
	\begin{align}\label{eq_model_id1}
		&{h}(f) = \mathbf{s}_{RM}{(f)}\boldsymbol\Gamma(f)\mathbf{s}_{MT}{(f)}.
	\end{align} 
	Under the assumption of an ideal MTP,
	the terms $\mathbf{s}_{RM}{(f)}$ and $\mathbf{s}_{MT}{(f)}${ can be written as}
	$\mathbf{s}_{RM}{(f)} = \frac{\mathbf{Z}_{RM}{(f)}}{2Z_0}, 
	\mathbf{s}_{MT}{(f)} = \frac{\mathbf{Z}_{MT}{(f)}}{2Z_0}$ \cite{DR1,abrardo2023design}, {where} $\mathbf{Z}_{RM}{(f)}$ {and} $\mathbf{Z}_{MT}{(f)}$ {are the Z-parameter sub-matrices that relate the input currents to the output open voltages at the transmitter, receiver and MTP's ports. They correspond to the traditional link-gains of LOS communication channels.}  
	Given the far-field assumption, we consider the \ac{MTP} as being concentrated in a point, such as the origin of the $[\nu, \zeta]$ axes, and define a 3D space by introducing a third axis, $\eta$, (see Figure \ref{fig:assi}). A plane wave incident upon or reflected by the \ac{MTP} is characterized by the direction of arrival/departure, which corresponds to the line connecting the point $\mathbf{p}$, shown in Figure \ref{fig:assi}, and the origin of the axes. We define the elevation angle $\vartheta$ as the angle between the direction of the plane wave and the direction normal to the surface, i.e., the $\eta$ axis. The azimuth $\varphi$ is defined as the angle between the projection of the direction of the plane wave onto the $\nu-\zeta$ plane and the $\nu$ axis. These angles are defined in the intervals $\vartheta \in [-\pi/2, \pi/2]$ and $\varphi \in [0, 2\pi]$.\\ 
	We are now in the position of outlining the channel model for the scenario considering the response of the \ac{MTP} to a plane wave with elevation angle $\vartheta$ and azimuth angle $\varphi$. Starting from 
	the definition of the array response vector,
	\begin{equation}\label{array_vector_a}
		\begin{aligned}
			\mathbf{a}({\vartheta, \varphi,f}) & = \left[ a_{1}({\vartheta, \varphi,f}), \ldots, a_N({\vartheta, \varphi,f})  \right] \\&=  \left[e^{\jmath\textbf{k}^{\text{T}}({\vartheta, \varphi,f})\textbf{p}_1},\ldots,e^{\jmath\textbf{k}^{\text{T}}({\vartheta, \varphi,f})\textbf{p}_N}\right]^{\text{T}},
		\end{aligned}
	\end{equation} 
	where 
	\begin{equation}
		\label{eq:waveNumber}
		\textbf{k}({\vartheta, \varphi,f}) = \frac{2\pi}{\lambda}[\sin(\vartheta)\cos(\varphi),\sin(\vartheta)\sin(\varphi)]^{\text{T}} 
	\end{equation}
	$\lambda { = c/f}${, and $c$ being the speed of light, we note that }the terms $\mathbf{s}_{MT}$ and $\mathbf{s}_{RM}$ in this scenario depend on the incidence and reflection angles of the wave incident on the MTP. Indeed, denoting by $\Theta_{\text{inc}} = [\theta_{\text{inc}}, \phi_{\text{inc}}]$ and $\Theta = [\theta, \phi]$ the angle of arrival and departure expressed in compact form, we have 
	$\mathbf{s}_{MT}({\Theta_\text{inc}{, f}}) = g_{mt}(f)\mathbf{a}\left({{\Theta_\text{inc}, f}}\right)$, $\mathbf{s}_{RM}\left(\Theta{, f}\right)= g_{rm}(f)\mathbf{a}\left({\Theta, f}\right)$. 
	{$g_{mt}(f)$ and $g_{rm}(f)$ are the gains of the two links that depend on the BS-\ac{MTP} and \ac{MTP}-receiver distances, together with the radiation patterns of the antennas and of the MTP's single scattering element.
		For the sake of simplicity we may assume that in both the incidence and reflection plane at the \ac{MTP},
		the transmitter, receiver and all the \ac{MTP}'s elementary scatterers have an omnidirectional radiation pattern over the whole frequency bandwidth, thus the two terms are set to one.
		As an example, let us consider that the transmitting and receiving antennas and the \ac{MTP}'s elementary scatterers are modeled as identical $\zeta$-oriented metallic thin wire dipoles and lie in the $\nu\zeta$-plane. 
		Consequently, both the incidence and reflection plane lie in the  $\nu\eta-$plane, and the normalized radiation pattern of all the antennas and scatterers on that plane are independent by $\Theta_{\text{inc}}$ and $\Theta$.}\\
	Under these assumptions, we finally introduce the angle-frequency end-to-end channel gain for the \ac{BS}-\ac{MTP}-node channel as: \footnote{Note that, by considering a far-field model, we do not lose generality in using a SISO channel model like the one in \eqref{eq_model_1}, as single-stream MIMO transmission/reception systems can be easily incorporated into the gains $g_{rm}$ and $g_{mt}$ as a result of beamforming. The case of multi-stream MIMO is beyond the scope of this paper and is left to future research.}
	\begin{align}\label{eq_model_1}
		& {h}(\Theta,f) = \mathbf{s}_{RM}(\Theta{, f}) ~ \mathbf{\Gamma}(f) ~ \mathbf{s}_{MT}({\Theta_\text{inc}{, f}}).  
	\end{align}
	In this setting, it is easy to show that the \ac{MTP} provides the maximum scattered energy towards the angle $\Theta(f)$ provided that, at each of its scattering elements, the following phase shift is applied to the impinging signal 
	\begin{equation}
		\begin{aligned}
			\label{eq:phase_analytical_thetak}
			\Pn(\Theta_{\text{inc}},& {f}) = -\angle{{a}_n\left({\Theta_\text{inc}, f}\right)} - \angle{a_n\left({\Theta(f), f}\right)} \\
			&= - \frac{2 \pi \nu_n \Delta_{\nu}}{\lambda}\left(u_{\nu}(\Theta_{\text{inc}}) + u_{\nu}(\Theta(f))\right)  \\
			&\quad - \frac{2 \pi \zeta_n \Delta_{\zeta}}{\lambda}\left(u_{\zeta}(\Theta_{\text{inc}}) + u_{\zeta}(\Theta(f))\right)+\psi_0, 
		\end{aligned}
	\end{equation}
	where $\psi_0$ is a common phase offset, and from \eqref{array_vector_a}-\eqref{eq:waveNumber} $u_{\nu}(\Theta) = \sin(\theta)\cos(\phi)$, and $u_{\zeta}(\Theta) = \sin(\theta)\sin(\phi)$. 
	{In the following, under the assumption of  narrowband systems, we will consider a frequency-flat response for the terms $s_{RT}$, $\mathbf{s}_{MT}$, $\mathbf{s}_{RM}$, and $\mathbf{S}_{SS}$, thus neglecting the frequency dependence. This assumption is validated by the full wave simulator analysis reported in Section \ref{sec:Results}. Accordingly, the frequency selectivity properties, and therefore the \ac{MTP} effect, is obtained by optimizing the reflection coefficient $\boldsymbol{\Gamma}(f)$.}
	
	\section{Ideal MTP Design}\label{IDEAL_MTP}
	In this section, we define the characteristics of an ideal MTP, designed according to the channel model in \eqref{eq_model_1}.
	\subsection{\ac{MTP} Definition}
	
\begin{figure}
		\vspace{-\baselineskip}
		\centering
		\includegraphics[width =0.7\linewidth]{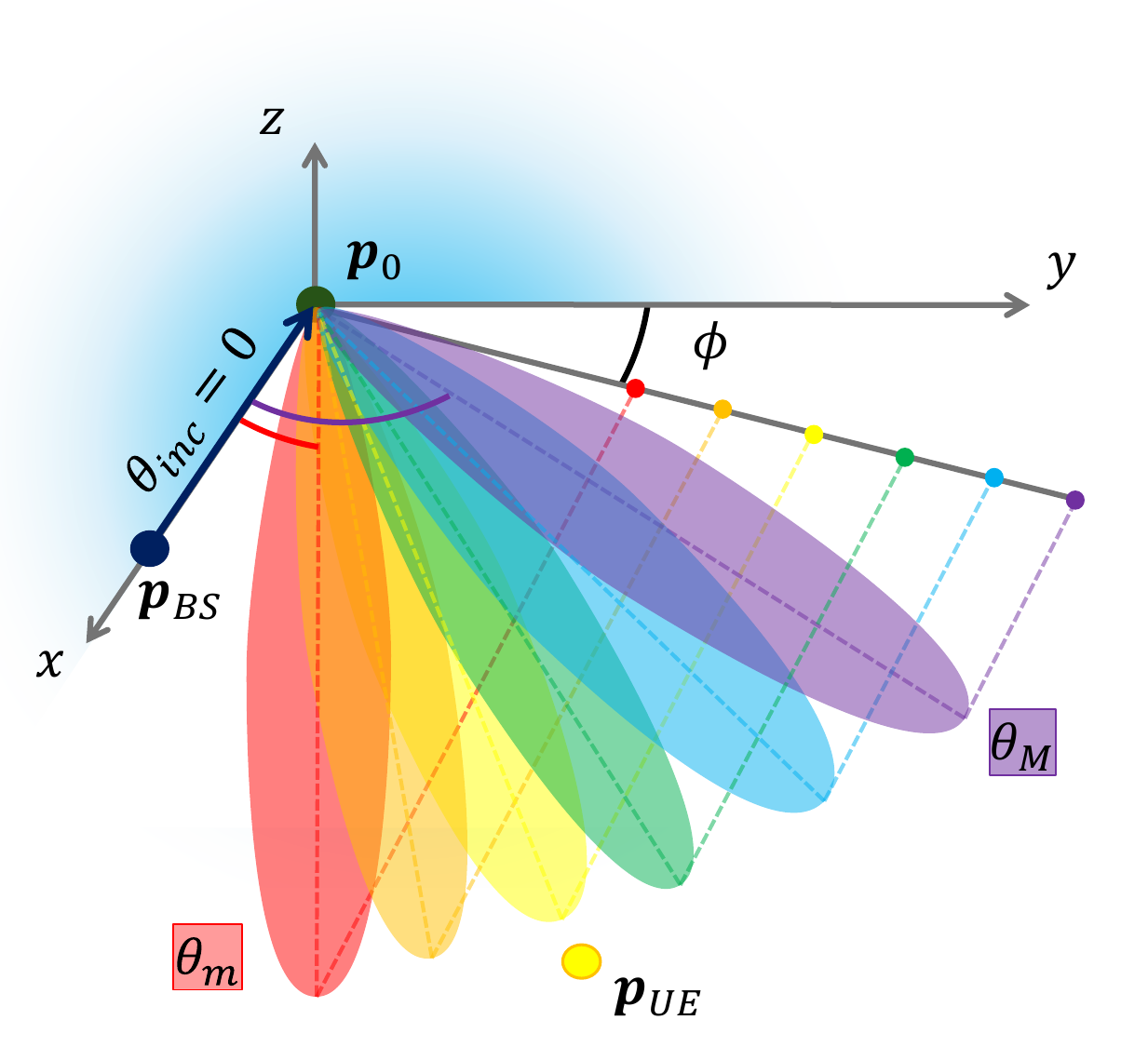}
		\caption{ \ac{MTP} in the elevational direction placed along the (y-z) axis, illuminating the angular range between $\theta_m$ and $\theta_M$}
		\label{fig:scenario}
		\vspace{-\baselineskip}
	\end{figure}

	\begin{definition}\label{MP_def}
		Given the carrier frequency $f_0$ and a bandwith $W$, an ideal \ac{MTP} in the elevation range is a frequency-selective metasurface in which, given the angle of incidence $\theta_\text{inc}$, the angle $\theta$ to which the beams are directed depends on the frequency, that is, $\theta(f) = \M(f)$, where $\M(f)$ is a monotonic function that maps the frequency range $\mathcal{F} = \left(f_0 - W/2, f_0 + W/2\right)$ to the elevation range $\mathcal{T} = \left(\theta_{m}, \theta_{M}\right)$, with $-\pi/2\le\theta_m\le\theta_M\le\pi/2$. 
	\end{definition}
	{A similar definition could be given for an \ac{MTP} defined with respect to the azimuth $\phi$ or jointly in the azimuth and elevation domains, as a mapping between $\mathbb{R}$ and $\mathbb{R}^2$. In a classic far-field scenario the coverage area that an \ac{MTP} must illuminate can be considered a plane, in the first approximation,  thus, defining the MTP only for one of the two angles is a situation of clear practical interest. In the following, we will consider only the \ac{MTP} defined in the elevation dimension, as further clarified in the graphical representation of an MTP reported in Figure \ref{fig:scenario}.
		In this case, we assume, without loss of generality, $(\nu, \zeta) = (y, z)$, and we show various beams assuming $\phi$ fixed. With an incident signal direction of $\theta_\text{inc} = 0$, the \ac{MTP} creates a continuous set of beams, represented in different colors, within the angular range $\left(\theta_{m}, \theta_{M}\right)$: this corresponds to illuminate an area in front of the \ac{MTP}, a situation of relevant practical interest.}
	It is easy to verify that the \ac{MTP} proposed in \cite{dardari2021using} corresponds to the general definition given in Definition \ref{MP_def} for a mapping function $\M(f)$ defined as:
	\begin{equation}
		\begin{aligned}
			\label{eq:Meta_Davide}
			\M(f) = \sin^{-1}\left(\alpha(f-f_0)+\gamma\right),
		\end{aligned}
	\end{equation}
	that is, the terms $u_{\nu}(\Theta) = \sin(\theta)\cos(\phi)$ and $u_{\zeta}(\Theta) = \sin(\theta)\sin(\phi)$ introduced in \eqref{eq:phase_analytical_thetak}, are linear functions of $f$. 
	The choice adopted in \eqref{eq:Meta_Davide} is just one of the possible mappings that define an \ac{MTP}. As a matter of fact, one could design the mapping as
	$\M(f) = \theta_m + \frac{2\left(f-f_0\right)}{W}\left(\theta_M-\theta_m\right)$, 
	corresponding to the uniform mapping case in which each frequency variation produces a proportional change in the angle. 
	{This choice, however, would produce a non-linear phase with frequency, whereas the mapping in \eqref{eq:Meta_Davide} has the convenient property of featuring a linear phase which, as we will see, allows the realization of an MTP with a frequency response that has the same shape for all possible angles $\Theta$.} 
	This means that if the MTP is used to serve nodes located at different angles $\Theta$, they will all be characterized by the same bandwidth. Additionally, the choice of a linear phase makes easier the practical implementation with standard circuitry, as will be shown in the second part of this paper.
	
	To elaborate, enforcing the general definition of the \ac{MTP}, $\alpha$ and $\gamma$ can be set in order to obtain the desired behavior, i.e., $\M(f_0-W/2) = \theta_m$ and $\M(f_0+W/2) = \theta_M$, which yields 
	\begin{equation}
		\label{eq:Meta_Davide2}
		\alpha =\frac{\sin(\theta_{M})-\sin(\theta_{m})}{W},\quad
		\gamma = \frac{\sin(\theta_{M})+\sin(\theta_{m})}{2}.
	\end{equation}
	Therefore,  
	the design choice in \eqref{eq:Meta_Davide2} implies that the phase of the reflection coefficient of the generic $n$-th element of the \ac{MTP} becomes
	{
		\begin{align}
			\label{eq:reflectionCoeffOurModel}
			\Pn(&\Theta_{\text{inc}}{, f}) = \\
			& =- \frac{2 \pi \nu_n \Delta_{\nu}}{\lambda}\left(u_{\nu}(\Theta_{\text{inc}}) +  \left(\alpha(f-f_0)+\gamma\right) \cos(\phi)\right) \nonumber\\
			& \, \, - \frac{2 \pi \zeta_n \Delta_{\zeta}}{\lambda}\left(u_{\zeta}(\Theta_{\text{inc}}) +  \left(\alpha(f-f_0)+\gamma\right) \sin(\phi)\right)+\psi_0. \nonumber
		\end{align}
	}
	
	\begin{figure}
		\vspace{-\baselineskip}
		\centering
		\includegraphics[width = 0.7\linewidth]{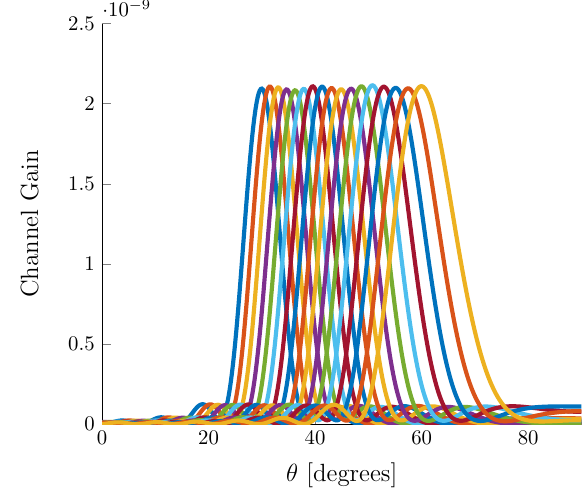}
		
		\caption{Channel gain for the ideal \ac{MTP} implementation as in \eqref{eq:Meta_Davide}, for $\Delta_y = \lambda_0/2$.} 
		\label{fig:idealGain}
	\end{figure}
	Figure \ref{fig:idealGain} shows an example of a radiation pattern obtained considering the parameters indicated in Table \ref{tab:parameters}, where $\lambda_0 = c/f_0$, $D_b$ and $D_u$ are the distances of the BS and the node from the origin of the reference system, respectively. 
	{Hereafter, as in \cite{abrardo2023design}, it is  considered that the transmitting and receiving antennas and the \ac{MTP} elementary scatterers are modeled as identical $\zeta$-oriented metallic thin wire dipoles with a radius $\lambda_0/500$ and length $0.46 \lambda_0$.\footnote{The dipole length  has been chosen to have nearly resonant elements characterized by a low reactance of the input impedance. As mentioned above, with this choice we can assume $g_{mt}=g_{rm}=1$.}} The figure illustrates the channel gain $\left|h(\Theta,f_k)\right|^2$ as a function of the elevation $\theta$. This is depicted for a discrete set of frequencies $\mathcal{F}_K = \{f_k\} \subseteq \mathcal{F}$, for $k = 1, \ldots, K$, obtained by uniformly sampling the frequency interval $\mathcal{F}$ and for $\phi = 0$. 
	According to the mapping defined in \eqref{eq:Meta_Davide} and \eqref{eq:Meta_Davide2} and from \eqref{eq:reflectionCoeffOurModel}, we get that for each frequency $f_k$ the maximum is obtained at the angles $\theta_k = \mathcal{M}(f_k)$.
	Note that due to the non-linear mapping in \eqref{eq:Meta_Davide}, the angles are not uniform, i.e., the beams tend to spread slightly as the angle increases. 
	\begin{table}
		\small
		\begin{center}
			\makebox[\linewidth]{
				\begin{tabular}{ |c|c|} 
					\hline
					Parameter & Value\\
					\hline
					$f_0$ &  $3.6 GHz$ \\ \hline
					$W$ &  $100 MHz$ \\ \hline
					$\tinc$ &  $0$ \\ \hline
					$\mathcal{T}$ & $[\pi/6, \pi/3]$ \\ \hline
					$\phi$ &  $0$ \\ \hline
					$\textbf{p}_M$ & $[0, 0, 0]$ \\ \hline
					$(\nu, \zeta)$ & $(y, z)$ \\ \hline
					$\Delta y$ & $\frac{\lambda_0}{2}$ \\ \hline
					$\Delta z$ & $\frac{3\lambda_0}{4}$ \\ \hline
					$I$ &  $16$ \\ \hline
					$J$ &  $4$ \\ \hline
					$\textbf{p}_{BS}$ & $[D_b(\cos(\tinc), \sin(\tinc), 0)]$ \\ \hline
					$D_b$ & $10$m \\ \hline 
					$\textbf{p}_{UE}(\theta_k)$ & $[D_u(\cos(\theta_k), \sin(\theta_k), 0)]$ \\ \hline
					$D_u$ & $20$m \\ \hline 
				\end{tabular}
			}
		\end{center}
		\caption{System parameters}
		\label{tab:parameters}
		\vspace{-\baselineskip}
	\end{table}
	
	\subsection{Available Bandwidth}
	\label{sec:AvailableBW}
\begin{figure}
		\vspace{-\baselineskip}
		\centering
		\includegraphics[width = 0.7\linewidth]{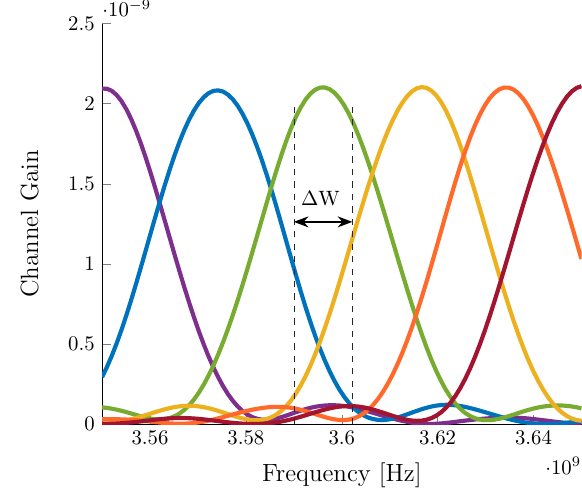}
		\caption{Channel gain for $f \in \mathcal{F}$ for the ideal \ac{MTP} implementation as in \eqref{eq:Meta_Davide}, for $\Delta_y = \lambda_0/2$.}
		\label{fig:ChGainIdFreq}
		\vspace{-\baselineskip}
	\end{figure}

\begin{figure}
		\centering
		\includegraphics[width=0.9\linewidth]{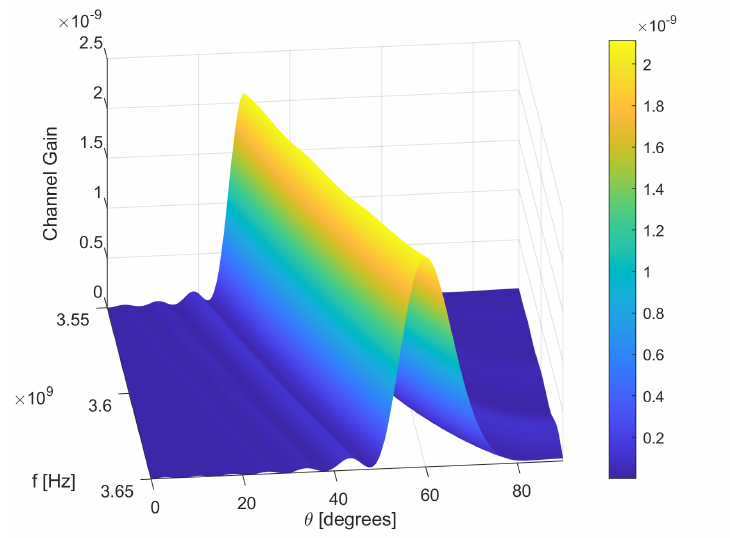}
		\caption{Channel gain for the ideal $\Delta_y = \lambda_0/2$ \ac{MTP} implementation as in \eqref{eq:Meta_Davide}, for $f \in \mathcal{F}$ and $\theta \in \mathcal{T}$.}
		\label{fig:3DID}
		\vspace{-\baselineskip}
	\end{figure}

	Let's now denote $h_M(k) = h([\theta_k,\phi], f_k)$ the maximum of the beams obtained for each frequency $f_k$. 
	{ For a generic frequency shift $\Delta f$ and from the }mapping rule in \eqref{eq:Meta_Davide} we have
	$\sin(\mathcal{M}(f+\Delta f)) = \sin(\mathcal{M}(f)) + \alpha \Delta f$.
	Then, considering a generic frequency $f_k$ and the angle $\theta_k = \mathcal{M}(f_k)$, it is easy to derive
	\begin{equation}
		\label{eq_BW3}
		\begin{aligned}
			& h\left([\theta_k,\phi],f_k+\Delta f\right) \\
			& = \frac{h_M(k)}{N} \sum\limits_{n=1}^{N}e^{-j\frac{2 \pi \alpha\left(\nu_n \Delta_{\nu}\cos(\phi)+\zeta_n \Delta_{\zeta}\sin(\phi)\right)\Delta f}{\lambda_0}}.
		\end{aligned}
	\end{equation}
	Elaborating from \eqref{eq_BW3}, we have
	\begin{equation}
		\label{eq_BW4}
		\begin{aligned}
			& h\left([\theta_k,\phi],f_k+\Delta f\right)  = 
			\\ & = \frac{h_M(k)}{N} \frac{\sin\left(\frac{\pi I \alpha \Delta_{\nu}\Delta f \cos(\phi)}{\lambda_0}\right)}{\sin\left(\frac{\pi \alpha \Delta_{\nu}\Delta f \cos(\phi)}{\lambda_0}\right)} \frac{\sin\left(\frac{\pi J \alpha \Delta_{\zeta}\Delta f \sin(\phi)}{\lambda_0}\right)}{\sin\left(\frac{\pi \alpha \Delta_{\zeta}\Delta f \sin(\phi)}{\lambda_0}\right)},
		\end{aligned}
	\end{equation}
	i.e., the frequency response is composed of a lobe centered at $f_k$ with the first null at $\Delta f = \min\left(\frac{\lambda_0}{I \alpha \Delta \nu \cos(\phi)},\frac{\lambda_0}{J \alpha \Delta \zeta \sin(\phi)}\right)$. Note that the normalized channel gain $\left|h\left(\theta_k,f_k+\Delta f\right)\right| /h_M(k)$ does not depend on $f_k$ and, consequently,
	on the elevation angle at which the beam is created. As a result, thanks to the mapping in \eqref{eq:Meta_Davide}, the bandwidth of the MTP in the elevation range does not depend on the elevation. Therefore, we define the bandwidth $\Delta W$ as the frequency range within which the channel response relative to a certain angle does not fall below a fraction $1-\omega$ of its maximum value, namely: 
	$\Delta W = \left\{ 2\Delta f : \left| h\left([\theta_k,\phi],f_k+\Delta f\right)\right|^2 = (1-\omega)^2\left|h_M(k)\right|^2\right\}$.
	For example, if we consider the case in which the geometry of the problem is almost planar, that is, the signal incident on the MTP arrives from a direction roughly belonging to the plane normal to the MTP, we have $\phi \approx 0$. In this case, the third fractional term of \eqref{eq_BW4} can be approximated to $J$ such that, from \eqref{eq_BW3} and \eqref{eq:Meta_Davide2}, for small $\omega$, it is possible to derive an approximate calculation of the bandwidth performing a third-order Taylor series expansion of the numerator in \eqref{eq_BW4}. With simple steps, we obtain
	$\Delta W \approx \sqrt{6\omega}\left(\frac{2W \lambda_0}{\pi I \Delta \nu \left(\sin(\theta_M) - \sin(\theta_m)\right)}\right)$.
	Considering, for example, the case in Figure \ref{fig:idealGain}, we have a bandwidth $\Delta W \approx 12$ MHz for $\omega = 0.05$. It is worth noting that the bandwidth decreases as the size $I \Delta_\nu$ of the MTP increases: as expected, a larger MTP results in higher gains but smaller bandwidths. Finally, the bandwidth is inversely proportional to the angular spreading that the MTP aims to achieve.
	{Figure \ref{fig:ChGainIdFreq}  reports a graphical representation of the frequency response for different frequencies $f_k$, or, similarly, different angles $\theta_k$, referring to the same scenario as in Figure \ref{fig:idealGain}. For ease of reading, only a subset of the frequencies considered in Figure \ref{fig:idealGain} is shown. Note that the bandwidth is the same for all the frequencies/angles considered. The figure also highlights the bandwidth resolution $\Delta W$ for $\omega = 0.05$, corresponding to the frequency range within which the square of the frequency response remains within a fraction of $(1-0.95)^2 \approx 0.9$ times its maximum value.
		Finally, Figure \ref{fig:3DID} provides a three-dimensional view of the response $h(\Theta, f)$ for the case shown in Figure \ref{fig:idealGain}}. 
	
	\subsection{MTP in the Presence of Multipath}
	{In addition to the \ac{LOS} component considered in \eqref{eq_model_1}, in environments with natural scattering, both the BS-MTP and MTP-receiver links may experience \ac{NLOS} components due to multipath. In the following an analysis of the multipath effects on MTP is presented. First of all, it is worth noting that the BS-MTP link can be assumed to have a strong LOS signal, as the two nodes can be positioned optimally to minimize scattering, thus, only the multipath in the MTP-receiver link is considered below. The analysis however, is reciprocal, meaning that the case of multipath in the BS-MTP link could be derived similarly. Dual multipath effects are ignored since their impact on the received signal is negligible.}
	{For the derivation a central frequency $f_k$ and the corresponding angle $\Theta_k$ are considered, with the assumption that $\Delta W$ is smaller than the coherence bandwidth of the channel, i.e., a flat fading model can be used for the multipath component. A generalized spatially correlated Rician fading model characterizes the total MTP-receiver channel  $\bar{\mathbf{s}}_{RM}$ as \cite{1499046}}
	\begin{equation}
		\begin{aligned}
			\label{eq_MP1}
			\bar{\mathbf{s}}_{RM} = \sqrt{\frac{\kappa_R}{\kappa_R+1}} \mathbf{s}_{RM}({\Theta_k}) + \sqrt{\frac{1}{\kappa_R+1}} g_{rm} \tilde{\mathbf{s}}_{RM}
		\end{aligned},
	\end{equation} 
	where $\kappa_R$ is the Rician factor and $\tilde{\mathbf{s}}_{RM}$ is the Rayleigh multipath component, characterized by a given spatial correlation. 
	Referring to spatially correlated multipath models considered in \cite{Bjornson2021Rayleigh,Demir2022Channel}, and simplifying the analysis by setting $\phi = 0$, we can write $\tilde{\mathbf{s}}_{RM} = \int\nolimits_{-\pi/2}^{\pi/2} s(\theta) z(\theta) \mathbf{a}\left({\theta}\right) d\theta$   
	where $z(\theta) \sim \mathcal{CN}(0, 1)$. $\mathbb{E}[z(\theta) z^*(\theta')] = \delta(\theta - \theta')$ is the uncorrelated fading term while $s(\theta)$ is the normalized gain for each angular direction, with $s^2(\theta)$ corresponding the spatial scattering distribution that satisfies $\int\nolimits_{-\pi/2}^{\pi/2} s^2(\theta) d\theta = 1$. It is then possible to derive the total channel response as:
	\begin{align}
		\label{eq_MP3}
		& \bar{h}([\theta_k,0],f_k)  = \bar{h}(\theta_k,f_k) = \bar{\mathbf{s}}_{RM} ~ \mathbf{\Gamma}(f) ~ \mathbf{s}_{MT}({\Theta_\text{inc}}) \\ & = \sqrt{\frac{\kappa_R}{\kappa_R+1}}h_M(k) + \sqrt{\frac{1}{\kappa_R+1}} \int\limits_{-\pi/2}^{\pi/2} s(\theta) z(\theta) {h}(\theta,f_k) d\theta .\nonumber
	\end{align}
	Let's now introduce the effective Rician factor $\kappa_{\text{eff}}$ as the ratio between the LOS power and the mean square value of the second term in \eqref{eq_MP3}:
	\begin{align}
		\label{eq_MP4}
		\kappa_{\text{eff}}  & = \frac{\kappa_R\left|h_M(k)\right|^2}{\int\limits_{-\pi/2}^{\pi/2}\int\limits_{-\pi/2}^{\pi/2} s(\theta)s(\theta')h(\theta,f_k)h^*(\theta',f_k)\mathbb{E}[z(\theta) z^*(\theta')]d\theta d\theta'}\nonumber \\ & = \frac{\kappa_R} {\displaystyle\int\nolimits_{-\pi/2}^{\pi/2} s^2(\theta) \frac{\left|{h}(\theta,f_k)\right|^2}{\left|h_M(k)\right|^2} d\theta}.
	\end{align}
	Since $\left|{h}(\theta,f_k)\right| \le \left|h_M(k)\right|$, in general we have $\kappa_{\text{eff}} \ge \ \kappa_R$. 
	More specifically, the MTP can effectively filter and significantly reduce the impact of multipath beyond the Rician factor $\kappa_R$ when the multipath exhibits wide angular spreading. As an example, in the case of isotropic multipath where $s(\theta) = \frac{1}{2\pi}$, typical for sub-6 GHz frequencies, the shape of $h(\theta, f_k)$ (as shown in Figure \ref{fig:idealGain}) leads to $\kappa_{\text{eff}} >> \kappa_R$. This results in the multipath effect being negligible. On the other hand, if the multipath had narrow angular spreading around $\theta_k$, the effect could not be ignored but would simply add to the direct component without significantly affecting the MTP design criteria.

	\section{LC Circuit Design}
	\label{sec:CircuitModelDesign}
	In the following we analyze the problem of designing a circuit model capable of implementing a frequency-selective metasurface with the mapping in \eqref{eq:Meta_Davide}.
	To embody the desired phase-shift behavior of an \ac{MTP}, from \eqref{reflection_coeff}, we have
	\begin{equation}
		\label{eq:phase_reale}
		\Pn(f) = \measuredangle{\Gn} = - 2 \arctan \frac{\Xn(f)}{Z_0} + \pi.
	\end{equation}
	By appropriately adjusting the reactance $\Xn$, it is possible to achieve any phase shift and thus to steer the beams in the desired direction according to \eqref{eq:phase_analytical_thetak}.
	
	\subsection{Circuit Model Analysis}
	Without loss of generality, we consider the elevation $\phi = 0$ in the following so that $\Theta_\text{inc} = (\theta_\text{inc},0)$, $\Theta = (\theta,0)$, and the second term of the phase in \eqref{eq:reflectionCoeffOurModel} is null. Accordingly, we can consider a single line of the \ac{MTP}, i.e., $n = \nu_n = 1, \ldots, I$. From \eqref{eq:reflectionCoeffOurModel} and \eqref{eq:phase_reale}, to realize the frequency selective response of the \ac{MTP} is necessary that
	\begin{equation}
		\label{eq:impedanceTan}
		\begin{aligned}
			\Xn(f) = & 
			Z_0 \tan\left[\frac{\pi \nu_n \Delta_\nu}{\lambda}\left(\sin(\theta_\text{inc})+\alpha(f-f_0)+\gamma\right)+\pi/2\right]
		\end{aligned}
	\end{equation}
	The relationship in \eqref{eq:impedanceTan} allows for deriving an important property of the impedance circuits that characterize the unit cells of the MTP.
	
	\begin{lemma}
		\label{lemma1}
		In order to implement the \ac{MTP} functionalities for a sufficiently large metasurface, the ideal reactance in the bandwidth of interest, assuming $W << f_0$, is characterized by a number of resonant and anti-resonant frequencies that alternate with each other.  
	\end{lemma}
	\begin{figure}
		\vspace{-\baselineskip}
		\centering
%
%
\definecolor{mycolor1}{rgb}{0.00000,0.44706,0.74118}%
\definecolor{mycolor2}{rgb}{0.85098,0.32549,0.09804}%
\begin{tikzpicture}

\begin{axis}[%
width=0.7\linewidth,
xmin=3550000000,
xmax=3650000000,
xlabel style={font=\color{white!15!black}},
xlabel={Frequency [Hz]},
ymin=-200,
ymax=20,
ylabel style={font=\color{white!15!black}},
ylabel={$\text{X}_\text{n}\text{(f)}$},
axis background/.style={fill=white},
xticklabel style={/pgf/number format/fixed,/pgf/number format/precision=3}, 
axis x line*=bottom,
axis y line*=left,
xmajorgrids,
ymajorgrids,
legend style={legend cell align=left, align=left, draw=white!15!black}
]
\addplot [color=mycolor1, line width=2.0pt]
  table[row sep=crcr]{%
3550000000	-28.8675134594813\\
3550781250	-28.2716801680692\\
3551562500	-27.681869679642\\
3552343750	-27.0978973510123\\
3553125000	-26.5195840429821\\
3553906250	-25.9467558812844\\
3554687500	-25.3792440288894\\
3555468750	-24.8168844690219\\
3556250000	-24.2595177982753\\
3557031250	-23.7069890292517\\
3557812500	-23.159147402187\\
3558593750	-22.6158462050588\\
3559375000	-22.0769426017019\\
3560156250	-21.5422974674883\\
3560937500	-21.0117752321519\\
3561718750	-20.4852437293664\\
3562500000	-19.9625740527061\\
3563281250	-19.4436404176416\\
3564062500	-18.9283200292422\\
3564843750	-18.4164929552766\\
3565625000	-17.9080420044198\\
3566406250	-17.4028526092913\\
3567187500	-16.9008127140655\\
3567968750	-16.4018126664072\\
3568750000	-15.905745113502\\
3569531250	-15.4125049019606\\
3570312500	-14.9219889813889\\
3571093750	-14.4340963114288\\
3571875000	-13.9487277720799\\
3572656250	-13.4657860771269\\
3573437500	-12.9851756905036\\
3574218750	-12.506802745433\\
3575000000	-12.0305749661921\\
3575781250	-11.5564015923555\\
3576562500	-11.0841933053802\\
3577343750	-10.6138621573992\\
3578125000	-10.145321502099\\
3578906250	-9.67848592755971\\
3579687500	-9.21327119094324\\
3580468750	-8.74959415491895\\
3581250000	-8.28737272572168\\
3582031250	-7.82652579273959\\
3582812500	-7.36697316953524\\
3583593750	-6.90863553620534\\
3584375000	-6.45143438298865\\
3585156250	-5.99529195503455\\
3585937500	-5.54013119824794\\
3586718750	-5.08587570612817\\
3587500000	-4.63244966752253\\
3588281250	-4.17977781521763\\
3589062500	-3.72778537529235\\
3589843750	-3.27639801716011\\
3590625000	-2.82554180422806\\
3591406250	-2.37514314510307\\
3592187500	-1.92512874527562\\
3592968750	-1.47542555921393\\
3593750000	-1.02596074280131\\
3594531250	-0.576661606050604\\
3595312500	-0.127455566030908\\
3596093750	0.321729900058957\\
3596875000	0.770967301734372\\
3597656250	1.22032918205043\\
3598437500	1.66988816447369\\
3599218750	2.11971699986432\\
3600000000	2.5698886136331\\
3600781250	3.02047615313887\\
3601562500	3.47155303539134\\
3602343750	3.92319299512628\\
3603125000	4.37547013331988\\
3603906250	4.82845896621065\\
3604687500	5.28223447489841\\
3605468750	5.73687215559136\\
3606250000	6.19244807057315\\
3607031250	6.64903889996474\\
3607812500	7.1067219943568\\
3608593750	7.56557542839075\\
3609375000	8.02567805536922\\
3610156250	8.48710956297892\\
3610937500	8.94995053021165\\
3611718750	9.41428248557209\\
3612500000	9.88018796666512\\
3613281250	10.3477505812572\\
3614062500	10.8170550699117\\
3614843750	11.2881873703014\\
3615625000	11.761234683306\\
3616406250	12.2362855410062\\
3617187500	12.7134298766923\\
3617968750	13.1927590970108\\
3618750000	13.6743661563745\\
3619531250	14.1583456337746\\
3620312500	14.6447938121325\\
3621093750	15.1338087603413\\
3621875000	15.6254904181514\\
3622656250	16.1199406840659\\
3623437500	16.6172635064145\\
3624218750	17.1175649777915\\
3625000000	17.6209534330452\\
3625781250	18.1275395510238\\
3626562500	18.6374364602875\\
3627343750	19.1507598490153\\
3628125000	19.6676280793401\\
3628906250	20.1881623063682\\
3629687500	20.7124866021457\\
3630468750	21.2407280848556\\
3631250000	21.773017053545\\
3632031250	22.3094871286977\\
3632812500	22.8502753989903\\
3633593750	23.3955225745884\\
3634375000	23.9453731473619\\
3635156250	24.4999755584242\\
3635937500	25.0594823734239\\
3636718750	25.6240504660453\\
3637500000	26.1938412102058\\
3638281250	26.769020681468\\
3639062500	27.3497598682194\\
3639843750	27.9362348932102\\
3640625000	28.5286272460799\\
3641406250	29.1271240275435\\
3642187500	29.7319182059598\\
3642968750	30.3432088870486\\
3643750000	30.9612015975824\\
3644531250	31.5861085839328\\
3645312500	32.2181491264182\\
3646093750	32.8575498704659\\
3646875000	33.504545175675\\
3647656250	34.1593774839498\\
3648437500	34.8222977079557\\
3649218750	35.4935656412488\\
3650000000	36.1734503915281\\
};
\addlegendentry{n = 1}

\addplot [color=mycolor2, line width=2.0pt]
  table[row sep=crcr]{%
3550000000	-186.602540378444\\
3550781250	-183.305065173903\\
3551562500	-180.114433084341\\
3552343750	-177.025411419381\\
3553125000	-174.033103500571\\
3553906250	-171.132922113193\\
3554687500	-168.320565435986\\
3555468750	-165.59199518308\\
3556250000	-162.943416724506\\
3557031250	-160.371260979468\\
3557812500	-157.872167900609\\
3558593750	-155.442971388553\\
3559375000	-153.080685494259\\
3560156250	-150.782491782731\\
3560937500	-148.545727745617\\
3561718750	-146.367876162505\\
3562500000	-144.246555321523\\
3563281250	-142.179510019331\\
3564062500	-140.164603269007\\
3564843750	-138.199808651686\\
3565625000	-136.283203254408\\
3566406250	-134.412961142402\\
3567187500	-132.587347319225\\
3567968750	-130.804712132723\\
3568750000	-129.06348608891\\
3569531250	-127.36217503946\\
3570312500	-125.699355711787\\
3571093750	-124.07367155361\\
3571875000	-122.48382886646\\
3572656250	-120.928593204986\\
3573437500	-119.406786020959\\
3574218750	-117.917281532818\\
3575000000	-116.459003803236\\
3575781250	-115.030924008785\\
3576562500	-113.632057887116\\
3577343750	-112.261463348328\\
3578125000	-110.918238238345\\
3578906250	-109.601518243131\\
3579687500	-108.310474923504\\
3580468750	-107.044313871147\\
3581250000	-105.802272977193\\
3582031250	-104.583620805437\\
3582812500	-103.387655062883\\
3583593750	-102.213701160889\\
3584375000	-101.061110860711\\
3585156250	-99.9292609977349\\
3585937500	-98.8175522791017\\
3586718750	-97.7254081498573\\
3587500000	-96.6522737231047\\
3588281250	-95.5976147699902\\
3589062500	-94.5609167656497\\
3589843750	-93.5416839875345\\
3590625000	-92.5394386627902\\
3591406250	-91.5537201616035\\
3592187500	-90.5840842336506\\
3592968750	-89.6301022849856\\
3593750000	-88.6913606928891\\
3594531250	-87.7674601563746\\
3595312500	-86.858015080205\\
3596093750	-85.9626529904195\\
3596875000	-85.0810139795048\\
3597656250	-84.2127501794739\\
3598437500	-83.3575252612247\\
3599218750	-82.5150139586647\\
3600000000	-81.6849016161823\\
3600781250	-80.8668837581395\\
3601562500	-80.0606656791461\\
3602343750	-79.2659620539536\\
3603125000	-78.4824965658816\\
3603906250	-77.7100015527573\\
3604687500	-76.9482176694132\\
3605468750	-76.1968935658446\\
3606250000	-75.455785580188\\
3607031250	-74.724657445728\\
3607812500	-74.0032800111912\\
3608593750	-73.2914309736297\\
3609375000	-72.5888946232369\\
3610156250	-71.8954615994788\\
3610937500	-71.2109286579597\\
3611718750	-70.5350984474743\\
3612500000	-69.8677792967311\\
3613281250	-69.2087850102608\\
3614062500	-68.5579346730514\\
3614843750	-67.9150524634774\\
3615625000	-67.2799674741155\\
3616406250	-66.6525135400609\\
3617187500	-66.0325290743809\\
3617968750	-65.4198569103613\\
3618750000	-64.8143441502204\\
3619531250	-64.2158420199848\\
3620312500	-63.6242057302331\\
3621093750	-63.0392943424358\\
3621875000	-62.4609706406278\\
3622656250	-61.889101008168\\
3623437500	-61.323555309353\\
3624218750	-60.7642067756612\\
3625000000	-60.2109318964183\\
3625781250	-59.6636103136855\\
3626562500	-59.12212472118\\
3627343750	-58.5863607670487\\
3628125000	-58.0562069603253\\
3628906250	-57.5315545809075\\
3629687500	-57.0122975929027\\
3630468750	-56.4983325611934\\
3631250000	-55.9895585710856\\
3632031250	-55.4858771509065\\
3632812500	-54.9871921974267\\
3633593750	-54.4934099039867\\
3634375000	-54.0044386912142\\
3635156250	-53.5201891402237\\
3635937500	-53.0405739281947\\
3636718750	-52.5655077662302\\
3637500000	-52.0949073394033\\
3638281250	-51.628691248899\\
3639062500	-51.1667799561698\\
3639843750	-50.7090957290216\\
3640625000	-50.255562589552\\
3641406250	-49.8061062638688\\
3642187500	-49.3606541335166\\
3642968750	-48.9191351885439\\
3643750000	-48.4814799821471\\
3644531250	-48.0476205868298\\
3645312500	-47.6174905520169\\
3646093750	-47.1910248630698\\
3646875000	-46.7681599016468\\
3647656250	-46.3488334073586\\
3648437500	-45.9329844406679\\
3649218750	-45.5205533469893\\
3650000000	-45.1114817219403\\
};
\addlegendentry{n = 2}

\end{axis}
\end{tikzpicture}%
		\input{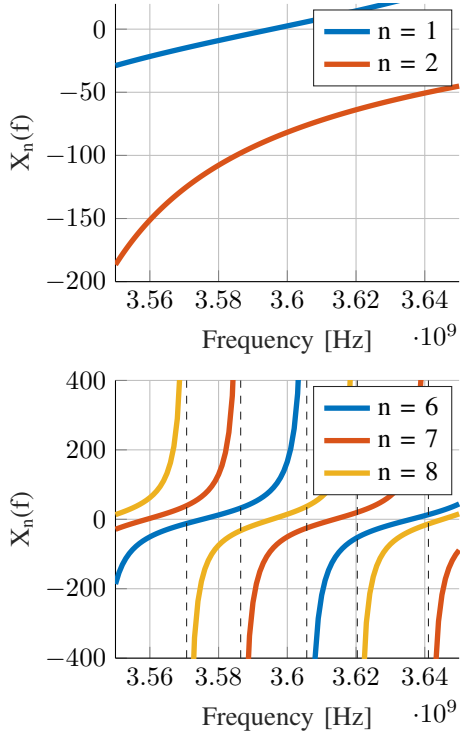}
		\caption{Ideal frequency dependent reactance for the proposed ideal reflection design according to the system parameters of Table \ref{tab:parameters}.}
		\label{fig:idealReactance}
		\vspace{-\baselineskip}
	\end{figure}
	\begin{proof} 
		The proof for Lemma \ref{lemma1} comes from  \eqref{eq:impedanceTan}. Since the tangent function has poles occurring at intervals of $\pi$, and considering the mapping property of the MTP, the argument of the tangent in \eqref{eq:impedanceTan} spans an interval of $\frac{\pi \nu_n \Delta_\nu}{\lambda_M}\sin(\theta_M)-\frac{\pi \nu_n \Delta_\nu}{\lambda_m}\sin(\theta_m) \approx \frac{\pi \nu_n \Delta_\nu}{\lambda_0}\left(\sin(\theta_M)-\sin(\theta_m)\right)$ over the frequency range $\mathcal{F}$. Therefore, regardless of the specific mapping $\mathcal{M}(f)$, $\Xn(f)$ must necessarily exhibit at least  $\Poln$ poles within $\mathcal{F}$ where 
		\begin{equation}
			\label{eq:poles}
			\Poln = \left\lfloor 
			\frac{\nu_n \Delta_\nu}{\lambda_0}\left(\sin(\theta_M)-\sin(\theta_m)\right)\right\rfloor.
		\end{equation}
		In other words, for angular spreads that are not too small and for wide \acp{MTP}, circuits design must incorporate structures that implement multiple poles within the desired frequency band.
		This is visually demonstrated in Figure \ref{fig:idealReactance}, based on the $\M$-function in \eqref{eq:Meta_Davide}. The reactance is calculated considering a system with the parameters from Table \ref{tab:parameters} and with $Z_0 = 50$ $\Omega$. The figure shows that when $n$ increases, more resonant and anti-resonant frequencies appear, alternating with each other. 
	\end{proof}
	To design a circuit that satisfies Lemma \ref{lemma1}, one approach is to use a network of anti-resonant circuits composed of inductances and capacitances in parallel, as described in \cite{foster1924reactance}. This configuration, known as the Foster Circuit of the first form, enables a one-to-one correspondence between each parallel LC pair and each pole in the frequency band. While this network is not the only architecture capable of meeting the requirements of Lemma \ref{lemma1}, it is particularly noteworthy because, with suitable adjustments, it can achieve an almost linear phase response across frequencies. This feature will be explored more extensively  in later discussions.
	
	To further elaborate, the designed reactance $\bXn(f)$ of the Foster circuit can be computed at frequency $f$ as: 
	\begin{equation}
		\label{eq:real_reactance}
		\bXn(f) = \sum_{p = 1}^ {\Poln} \frac{2 \pi f \bLnp}{1- (2 \pi f)^2 \bLnp \bCnp} = \sum_{p = 1}^ {\Poln} \bXnp(f),
	\end{equation}
	where $\bLnp$, $\bCnp$, and $\bXnp(f)$ are the inductances, the capacitances, and the frequency-dependent reactances of $n$-th circuit to realize the pole in positions $f_{n,p} = \frac{1}{2\pi\sqrt{\bLnp \bCnp}}$, for $p = 1,2, \ldots, \Poln$.
	
	\subsection{Circuit Model Derivation}\label{Foster_opt}
	To obtain the pole at the desired frequency $f_{n,p}$ in \eqref{eq:real_reactance} we impose the condition
	\begin{equation}
		\label{eq:resonance}
		\bLnp \bCnp = \frac{1}{2\pi f_{n,p}},
	\end{equation}
	which allows \eqref{eq:real_reactance} to be expressed as a linear function of the inductances
	\begin{equation}
		\label{eq:real_reactanceLinform}
		\bXn(f) = \sum\nolimits_{p = 1}^ {\Poln} \betapn (f) \bLnp,
	\end{equation}
	where 
	\begin{equation}
		\label{eq:beta}
		\betapn(f) = \frac{2 \pi f}{1- (f/f_{n,p})^2}.
	\end{equation}
	
	{The calculation of the Foster circuit poles for each element of the \ac{MTP} can be easily performed, as reported in the proof of Lemma \ref{lemma1}, by considering that a pole must occur whenever the circuit reactance becomes infinite, i.e., whenever the phase of the reflection coefficient becomes a multiple of $2\pi$. From \eqref{eq:reflectionCoeffOurModel}, considering the poles of the circuit for the $n$-th element of the \ac{MTP}, $f_{n,p}$, we have
		\begin{equation}
			\label{eq:phase_designv2}
			{\Pn}(\tinc, f_{n,p}) = 2 \varkappa_{n,p} \pi.
		\end{equation}
		All the possible $\varkappa_{n,p}$, and hence, all the poles of the circuit, can be computed by considering the two extreme points $f = f_0 \pm W/2$. To elaborate, the possible $\varkappa_{n,p}$ are in the range $[\varkappa_{\text{min}}, \varkappa_{\text{max}}] \in \mathbb{Z}^{\Poln \times 1}$, with:
		\begin{align}
			\label{eq:kmin_kmax}
			&\varkappa_{\text{min}} = \big\lfloor-\frac{\nu_n \Delta_{\nu}}{\lambda} (\sin(\tinc)+\sin(\tm))\big\rfloor\\
			&\varkappa_{\text{max}} = \big\lceil-\frac{\nu_n \Delta_{\nu}}{\lambda}(\sin(\tinc)+\sin(\tM))\big\rceil. \nonumber
		\end{align}
		Substituting the possible $\varkappa_{n,p}$ into \eqref{eq:phase_designv2} all the $f_{n,p}$ are computed, for $p = 1, \ldots, \Poln$ as
		\begin{equation}
			\label{eq:poles}
			f_{n,p} = -\frac{1}{\alpha}\left( \frac{\varkappa_{n,p} \lambda}{\nu_n\Delta_\nu}+\sin(\tinc)+\gamma\right)+ f_0.
		\end{equation}
	}
	
	By using the linear form introduced in \eqref{eq:real_reactanceLinform}, it is straightforward to employ the \ac{LS} method to determine the optimal values of inductances for each {circuit connected to each} scattering element. Specifically, denoting $\mathbf{\overline{l}_{n}} = \left(\bar{L}_{n,1},\ldots,\bar{L}_{n,\Poln}\right)$, and considering the discrete set of frequencies $\mathcal{F}_K$, we formulate the problem as follows: 
	\begin{align}
		\label{eq:LS}
		&\min_{\mathbf{\overline{l}_{n}}} \sum\nolimits_{k = 1}^K \left(\sum\nolimits_{p = 1}^{\Poln} \betapn(\fk)\bLnp - \Xn(\fk)\right)^2. 
	\end{align}
	
\begin{figure}
		\centering
		\includegraphics[width = 0.7\linewidth]{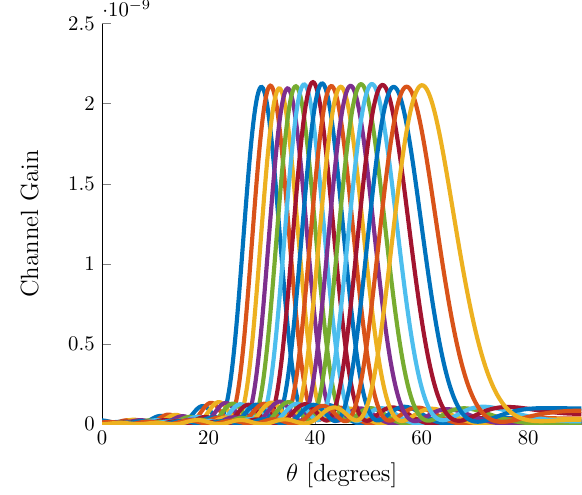}
		\caption{Realistic circuit design implementation by means of Foster's first form circuits, for $\Delta_y = \lambda_0/2$.}
		\label{fig:ChGainFosterReactanceIdealRIS}
		\vspace{-\baselineskip}
	\end{figure}

	{
		Problem \eqref{eq:LS} can be easily solved deriving the stationary points of the objective function with respect to every inductance variable. To elaborate, indicating with $g(\mathbf{\overline{l}_{n}})$ the objective function to be minimized and imposing ${\partial g(\mathbf{\overline{l}_{n}})}/{\partial_{\bar{L}_{n,i}}} = 0 $ $\forall i = 1, \ldots, \Poln$, we get
		\begin{equation}
			\label{eq:LS_gradient} 
			\sum_{k = 1}^K 2\left(\sum_{p = 1}^{P} \betapn(\fk)\bLnp - \Xn(\fk)\right) \beta_{n,i}(\fk) = 0.
		\end{equation}
		In matrix form we can express \eqref{eq:LS_gradient} as 
		\begin{equation}
			\label{eq:LS_gradient_mat}
			\mathbf{Q}_{n} \mathbf{\overline{l}_{n}} = \boldsymbol{\mu}_{n},
		\end{equation}
		where $\boldsymbol{\mu}_{n} \in \mathbb{R}^{[\Poln \times 1]^+}$ is the vector of positive known terms, with the $i$-th entry indicated as
		\begin{equation}
			\label{eq:knowterms}
			\mu_{n,i} = \sum\nolimits_{k = 1}^K \Xn(\fk) \beta_{n,i}(\fk),
		\end{equation}
		while the symmetric matrix $\mathbf{Q}_{n} \in \mathbb{R}^{[\Poln \times \Poln]}$ is positive semidefinite with entries
		\begin{equation}
			\label{eq:matrixcoeff}
			q_{n;i,p} = \sum\nolimits_{k = 1}^K 2 \beta_{n,i}(\fk) \betapn(\fk),  
		\end{equation}
		with $\{i,p\} \in \{1 \ldots, \Poln\}$. 
		{Due to its structure, the matrix $\mathbf{Q}_{n}$ is reasonably assumed full rank, as experimentally verified in all the considered cases. On the other hand, in the event this condition is not met, it will be sufficient to vary the set of discrete frequencies used for the fitting to obtain a full-rank matrix. Accordingly, } the optimal inductance solution is easily found as 
		\begin{equation}
			\label{eq:ComputedoptimalL}  \mathbf{\overline{l}_{n}}^\star = \mathbf{Q}_{n}^{-1} \boldsymbol{\mu}_{n}  
		\end{equation}
		and the capacitances are computed accordingly as
		\begin{equation}
			\label{eq:ComputedoptimalC}
			\bar{C}_{n,p}^\star = \frac{1}{\bar{L}_{n,p}^\star 2 \pi f_{n,p}}.
		\end{equation}
	} 
	The whole procedure is detailed in Algorithm \ref{alg:optimalLC}. 
	\begin{algorithm}
		\small
		\SetAlgoLined
		$\forall n =1, \ldots, N$\\
		-Compute $\varkappa_{\text{min}}$, $\varkappa_{\text{max}}$ \eqref{eq:kmin_kmax};\\
		-Set $\Poln$ \eqref{eq:poles} and compute $\alpha$, $\gamma$ \eqref{eq:Meta_Davide2};\\
		\For{$p = 1, \ldots, \Poln$}{
			-Derive $\varkappa_{n,p} \in [\varkappa_{\text{min}}, \varkappa_{\text{max}} ]$;\\
			-Compute the poles $f_{n,p}$ \eqref{eq:poles};\\
		}
		\For{$k = 1, \ldots, K$}{
			-Derive $\M(f_k)$, $\Pn(f_k)$ and $\Xn(f_k)$ \eqref{eq:Meta_Davide},\eqref{eq:reflectionCoeffOurModel}, \eqref{eq:impedanceTan};\\
			
			\For{$p = 1, \ldots, \Poln$}{
				-Compute the coefficients $\betapn(f_k)$ \eqref{eq:beta};\\
			}
		}
		
		\For{$i = 1, \ldots, \Poln$}{
			-Compute the known terms $\mu_{n,i}$ \eqref{eq:knowterms};\\ 
			\For{$p = 1, \ldots, \Poln$}{
				-Compute the coefficient matrix element $q_{n;i,p}$ \eqref{eq:matrixcoeff};\\
			}
		}
		-Compute the optimal inductances and capacitances $\bar{L}_{n,p}^\star$, $\bar{C}_{n,p}^\star$ (\eqref{eq:ComputedoptimalL},\eqref{eq:ComputedoptimalC};\\
		\caption{Foster circuit derivation}
		\label{alg:optimalLC}
	\end{algorithm}
	\begin{figure}
		
		\centering
%
%
\definecolor{mycolor1}{rgb}{0.30196,0.74510,0.93333}%
\definecolor{mycolor2}{rgb}{0.63529,0.07843,0.18431}%
\begin{tikzpicture}

\begin{axis}[%
width=0.7\linewidth,
unbounded coords=jump,
xmin=3550000000,
xmax=3650000000,
xlabel style={font=\color{white!15!black}},
xlabel={Frequency [Hz]},
ymin=-300,
ymax=300,
ylabel style={font=\color{white!15!black}},
ylabel={$\text{X}_\text{n}\text{(f)}$},
axis background/.style={fill=white},
axis x line*=bottom,
axis y line*=left,
xmajorgrids,
ymajorgrids,
legend style={legend cell align=left, align=left, draw=white!15!black},
xticklabel style={/pgf/number format/fixed,/pgf/number format/precision=3}, 
]
\addplot [color=mycolor1, line width=6.0pt]
  table[row sep=crcr]{%
3550000000	13.3974596215561\\
3551000000	18.1405642666362\\
3552000000	23.1908648592634\\
3553000000	28.6631235148837\\
3554000000	34.7061080894343\\
3555000000	41.5206993358079\\
3556000000	49.3894872479855\\
3557000000	58.7277861280808\\
3558000000	70.1770345100345\\
3559000000	84.7886707246649\\
3560000000	104.420853302792\\
3561000000	132.704684780234\\
3562000000	177.834381199899\\
3563000000	263.04336783257\\
3564000000	490.435939309049\\
3565000000	nan\\
3566000000	nan\\
3567000000	-315.586683122247\\
3568000000	-201.374728671445\\
3569000000	-146.158604665721\\
3570000000	-113.22139787786\\
3571000000	-91.0693463831931\\
3572000000	-74.944787134078\\
3573000000	-62.5199516622625\\
3574000000	-52.5194306622436\\
3575000000	-44.1840584501124\\
3576000000	-37.0319393253826\\
3577000000	-30.7406491109724\\
3578000000	-25.0844286459145\\
3579000000	-19.8984824917198\\
3580000000	-15.0575349381882\\
3581000000	-10.4622744826008\\
3582000000	-6.03032324601369\\
3583000000	-1.68984470096869\\
3584000000	2.62534213265137\\
3585000000	6.97988601112179\\
3586000000	11.4408259253246\\
3587000000	16.081827978595\\
3588000000	20.9883248163829\\
3589000000	26.2643644287838\\
3590000000	32.0424242835628\\
3591000000	38.4983178336828\\
3592000000	45.8750359647502\\
3593000000	54.5228958504455\\
3594000000	64.9710916809748\\
3595000000	78.0639965145563\\
3596000000	95.2431381177294\\
3597000000	119.196322416946\\
3598000000	155.586785827543\\
3599000000	218.764170002334\\
3600000000	358.819216703606\\
3601000000	nan\\
3602000000	nan\\
3603000000	-413.711923288468\\
3604000000	-238.65973039115\\
3605000000	-165.881441131268\\
3606000000	-125.54745809577\\
3607000000	-99.6020395579371\\
3608000000	-81.2805431658874\\
3609000000	-67.4744423996219\\
3610000000	-56.5533381852885\\
3611000000	-47.5778939838121\\
3612000000	-39.9669917913138\\
3613000000	-33.340047955471\\
3614000000	-27.4355388449053\\
3615000000	-22.065876171727\\
3616000000	-17.0909133075119\\
3617000000	-12.4015565704036\\
3618000000	-7.90909535640007\\
3619000000	-3.53785377469876\\
3620000000	0.780231235179978\\
3621000000	5.10998331489175\\
3622000000	9.51693194711703\\
3623000000	14.0713500701454\\
3624000000	18.8529258028159\\
3625000000	23.9567600563243\\
3626000000	29.5017169601481\\
3627000000	35.6428107911638\\
3628000000	42.5905919409612\\
3629000000	50.643070730886\\
3630000000	60.2411975611267\\
3631000000	72.0714227847791\\
3632000000	87.2700300881447\\
3633000000	107.870857276406\\
3634000000	137.918927577745\\
3635000000	186.787057941994\\
3636000000	282.281705585476\\
3637000000	559.889902131708\\
3638000000	nan\\
3639000000	-597.997143078898\\
3640000000	-291.872005224254\\
3641000000	-191.102110243627\\
3642000000	-140.388775608404\\
3643000000	-109.487560540996\\
3644000000	-88.4242520564762\\
3645000000	-72.947806093665\\
3646000000	-60.9383544101946\\
3647000000	-51.2185402827178\\
3648000000	-43.080309466775\\
3649000000	-36.0704877561879\\
3650000000	-29.8837492219575\\
};
\addlegendentry{$X_n(f)$ \eqref{eq:impedanceTan}}

\addplot [color=mycolor2, line width=2.0pt]
  table[row sep=crcr]{%
3550000000	8.50796238790048\\
3551000000	14.1431294077137\\
3552000000	19.9787673847823\\
3553000000	26.1435027418045\\
3554000000	32.7977323180682\\
3555000000	40.1521558823289\\
3556000000	48.4977021730262\\
3557000000	58.2568002367011\\
3558000000	70.0769767474909\\
3559000000	85.0148769846492\\
3560000000	104.933073700678\\
3561000000	133.466287870396\\
3562000000	178.811322741362\\
3563000000	264.202012673563\\
3564000000	491.733506205073\\
3565000000	nan\\
3566000000	nan\\
3567000000	-313.83313241759\\
3568000000	-199.552058542728\\
3569000000	-144.276329774176\\
3570000000	-111.291915601292\\
3571000000	-89.105134598762\\
3572000000	-72.9577400305758\\
3573000000	-60.5212033193747\\
3574000000	-50.5193270099562\\
3575000000	-42.1921811655931\\
3576000000	-35.0571484154183\\
3577000000	-28.7911312403476\\
3578000000	-23.1677458255678\\
3579000000	-18.0216188604166\\
3580000000	-13.2269407643693\\
3581000000	-8.68390717444672\\
3582000000	-4.30968549180497\\
3583000000	-0.0320200477430959\\
3584000000	4.21565582806048\\
3585000000	8.49834470800698\\
3586000000	12.8834084058631\\
3587000000	17.4448048241706\\
3588000000	22.2682260987345\\
3589000000	27.4579442913903\\
3590000000	33.146619614148\\
3591000000	39.5101965047829\\
3592000000	46.7917269443107\\
3593000000	55.3414873412522\\
3594000000	65.6884708257811\\
3595000000	78.6765762769457\\
3596000000	95.7463491452117\\
3597000000	119.583562087373\\
3598000000	155.846919210482\\
3599000000	218.874245412114\\
3600000000	358.714507075837\\
3601000000	nan\\
3602000000	nan\\
3603000000	-413.461076681282\\
3604000000	-238.65213255601\\
3605000000	-166.030643287495\\
3606000000	-125.825327679652\\
3607000000	-99.9953631669926\\
3608000000	-81.7815580621948\\
3609000000	-68.0777659519299\\
3610000000	-57.2547215387707\\
3611000000	-48.373634111953\\
3612000000	-40.8536211433893\\
3613000000	-34.3141547827162\\
3614000000	-28.4936541993502\\
3615000000	-23.2043969533458\\
3616000000	-18.3060455538793\\
3617000000	-13.6892702893526\\
3618000000	-9.26508582780051\\
3619000000	-4.95750619127991\\
3620000000	-0.698124163883842\\
3621000000	3.57826226040049\\
3622000000	7.93759635762614\\
3623000000	12.450602395256\\
3624000000	17.1974602616249\\
3625000000	22.2738067678569\\
3626000000	27.7990906059861\\
3627000000	33.9289647876842\\
3628000000	40.8746793480039\\
3629000000	48.9350135713928\\
3630000000	58.5517670660345\\
3631000000	70.412334780058\\
3632000000	85.6540635569107\\
3633000000	106.312013787942\\
3634000000	136.432677744288\\
3635000000	185.390824047215\\
3636000000	280.996236722013\\
3637000000	558.746631598288\\
3638000000	nan\\
3639000000	-598.905184178749\\
3640000000	-292.566349688246\\
3641000000	-191.56852778207\\
3642000000	-140.599776401234\\
3643000000	-109.410371329674\\
3644000000	-88.0217284190412\\
3645000000	-72.1782967091375\\
3646000000	-59.7552330865309\\
3647000000	-49.5695460566378\\
3648000000	-40.9067169457585\\
3649000000	-33.3060923030067\\
3650000000	-26.4536320966396\\
};
\addlegendentry{$\bXn(f)$ \eqref{eq:real_reactance}}

\end{axis}
\end{tikzpicture}%
		\caption{Comparison of the optimal and derived reactance for a single scattering element.}
		\label{fig:Xottimaederivatta}
		\vspace{-\baselineskip}
	\end{figure}
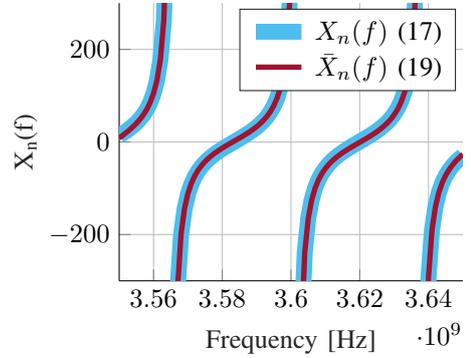
	For the proposed circuit design, the results are shown in Figures \ref{fig:Xottimaederivatta} and \ref{fig:ChGainFosterReactanceIdealRIS}. Figure \ref{fig:Xottimaederivatta} provides a graphical comparison between the optimal $X_n(f)$, calculated using \eqref{eq:impedanceTan}, and the one obtained from \eqref{eq:real_reactance} after optimization, demonstrating an almost perfect match. In Figure \ref{fig:ChGainFosterReactanceIdealRIS}, we illustrate the same example as in Figure \ref{fig:idealGain}, but with the \ac{MTP} implemented via the optimized Foster circuit. When compared with Figure \ref{fig:idealGain}, it is evident that the ideal \ac{MTP} effect can be almost ideally obtained using a real Foster's circuit implementation. 
	
	\section{MTP Design for a Realistic Multiport Network Model }
	\label{sec:RealisticSurfaceModels}
	\subsection{Effect of Model Mismatch}
	{As discussed in \cite{abrardo2023design}, the ideal conditions assumed in \eqref{eq_model_1} are not met for realistic metasurfaces.
		{{Indeed}, both structural scattering and mutual coupling among the elementary scatterers cannot be neglected {in the MTP's design}: the former may become significant when using loaded nearly-resonant elementary scatterers, and the latter may affect MTP performance when the scatterers densify. }}An interesting area of investigation could involve adopting technological solutions to minimize the effects of mutual coupling and mitigate structural scattering, thereby making model \eqref{eq_model_1} sufficiently adequate. This aspect is far beyond the scope of the present work {and is left for future investigations}. 
	
\begin{figure}
		\vspace{-\baselineskip}
		\centering
		\includegraphics[width = 0.7\linewidth]{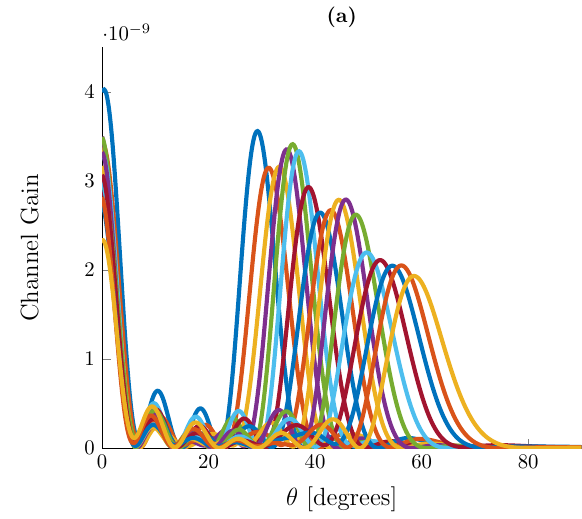}\\
		\includegraphics[width = 0.7\linewidth]{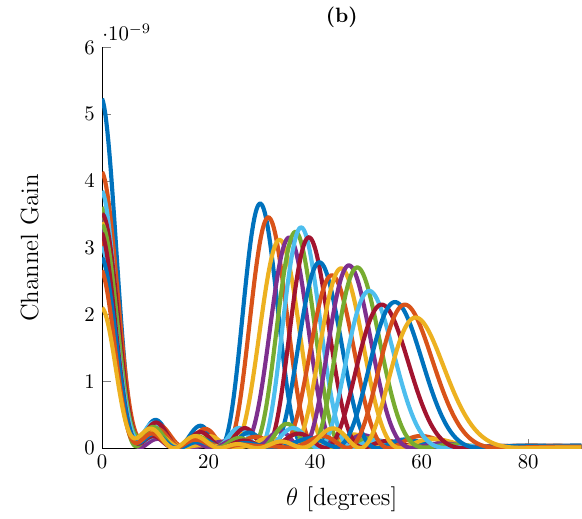}
		
		\caption{Channel gain with the Foster's first form circuit design and realistic \ac{MTP} model, for $\Delta_y = \lambda_0/2$ (a), and $\Delta_y = \lambda_0/4$ (b).  }
		\label{fig:ChGainFosterReactanceRealRISNoSpecular}
		\vspace{-\baselineskip}
	\end{figure}

	To illustrate the performance degradation due to model mismatch, we plug the reflection coefficients $\Gamma_n(f_k)$, obtained through the Foster circuit implementation shown in Algorithm \ref{alg:optimalLC} for the ideal phases in \eqref{eq:reflectionCoeffOurModel},
	into the multiport model in \eqref{eq_model_2} for any frequency $f_k \in \mathcal{F}_K$. 
	The results are presented in Figures \ref{fig:ChGainFosterReactanceRealRISNoSpecular}a and \ref{fig:ChGainFosterReactanceRealRISNoSpecular}b for inter-dipole distances of $\Delta_y = \lambda_0/2$ and $\Delta_y = \lambda_0/4$, respectively. 
	To ensure a fair comparison, the total \ac{MTP} size is kept constant, allowing for a higher number of elements with smaller spacing: 32 elements for $\Delta_y = \lambda_0/4$ and 16 for $\Delta_y = \lambda_0/2$. This configuration will be used in all subsequent analyses.\\
	Compared to Figures \ref{fig:idealGain} and \ref{fig:ChGainFosterReactanceIdealRIS}, a key difference is the appearance of multiple beams overlapping across all frequencies in the specular direction ($\theta_\text{inc} = 0^{\circ}$), a phenomenon not predicted by the ideal model. These overlapping beams cause slight interference with the desired beams, resulting in variations in beam intensity due to constructive or destructive interference with the specular component. Additionally, the beam gain decreases as the angle increases, an effect attributed to the mutual coupling between the elements. 
	Beyond the qualitative considerations mentioned, the most significant quantitative effect of model mismatch is a substantial reduction in system capacity, as will be shown in the results of Section \ref{sec:Results}.\\
	\textcolor{black}{Additionally, it should be noted that, as shown in \cite{Qian2021Mutual}, employing a smaller distance between \ac{MTP}'s elements is expected to improve channel gain performance, provided that an adequate model for the \ac{MTP} characterization is used during optimization. Conversely, neglecting mutual coupling or using an inadequate model can lead to a significant decline in performance. This is confirmed by the analysis of the previous figures, where the case $\Delta_y = \lambda_0/4$ shows significantly worse performance than the case $\Delta_y = \lambda_0/2$}. To address this issue, we propose an optimization approach tailored to the realistic model in \eqref{eq_model_2} in the following section.
	\subsection{Reflection Coefficient Optimization}
	\label{sec:ReflectionCoefficientOptimization}
	{In what follows, it is presented an approach for designing the reflection coefficient $\boldsymbol{\Gamma}$ aimed at optimizing the MTP behavior for the model given in \eqref{eq_model_2}. }
	This model exhibits nonlinear behavior in $\boldsymbol{\Gamma}$, making the optimization process particularly challenging. To tackle this issue, we will employ an \ac{AO} approach, leveraging the Sherman-Morrison inversion formula to decompose the problem into more manageable sub-problems. The solution obtained will also be subject to the constraint of maintaining phase linearity of the reflection coefficient with respect to frequency, ensuring the implementability with Foster circuits using the approach proposed in Section \ref{Foster_opt}.
	
	\subsubsection{Objective Function} 
	{Defining the set $\mathcal{T}_K = \left\{\theta_1,\ldots,\theta_K\right\} \in \mathcal{T}$, we consider a set of {$K = W/\Delta W$ users} at a distance $D_u$ from the \ac{MTP}, positioned at angles $\theta_k$ relative to the \ac{MTP}. For each of these users, we calculate the parameters of the model \eqref{eq_model_2}, which we denote as ${s}_{RT}^{(k)}$, $\mathbf{s}_{RM}^{(k)}$, $\mathbf{S}_{SS}^{(k)}$, and $\mathbf{s}_{MT}^{(k)}$. The channel gain for each user is:}
	\begin{equation}\label{model_RIS_MP}
		\tilde{h}^{(k)}= \mathbf{s}_{RT}^{(k)}+\mathbf{s}_{RM}^{(k)}\left(\boldsymbol{\Gamma}_k^{-1}-\mathbf{S}_{SS}^{(k)}\right)^{-1}\mathbf{s}_{MT}^{(k)},
	\end{equation}
	where $\boldsymbol{\Gamma}_k = \left\{e^{\jmath \psi_{k,n}}\right\}$, $\psi_{k,n}$ being the phase of the reflection coefficient of the $n$-th element of the \ac{MTP} at frequency $f_k = \mathcal{M}^{-1}(\theta_k)$.
	{The objective function of the problem is defined as the system's capacity under full load. To elaborate, let $P_t$ denote the transmitting power of the \ac{BS} and $N_0$ the noise power spectral density. Assuming for simplicity that the power is equally distributed among the users,\footnote{{While this is not the optimal strategy—ideally, a waterfilling approach would maximize performance—it serves as a reasonable approximation of the capacity since the channel gain is relatively uniform across the spectrum.}} each user receives $P = \frac{P_t}{K}$ }. {Given that the noise power for each user is $N_0 \Delta W$, the rate of this user is}
	\begin{equation}
		\begin{aligned}
			\label{eq_C1}
			R(f) = \Delta W \text{log}\left(1 + \frac{\left|\tilde{h}(k)\right|^2 P_t}{ W N_0}\right).
		\end{aligned}
	\end{equation}
	{Consequently the MTP system capacity can be derived as: }
	\begin{equation}
		\begin{aligned}
			\label{eq:objFunct}
			C_{{M}} & = \sum\nolimits_{k = 1}^ K \frac{W}{K} \text{log}\left(1+ \frac{\left|\tilde{h}(k)\right|^2 P_t}{ W 
				N_0}\right).
		\end{aligned}
	\end{equation}
	{ It is important to emphasize that the parameter $C_M$ in \eqref{eq:objFunct} is not intended to measure the actual sum-rate of a system with an MTP, as it refers to a scenario where all users are exclusively served by the MTP, with a large number of users uniformly distributed within its coverage area. {In practice, many users could be served directly by the BS via a direct path, which reduces the necessity of the MTP, while others might be served by other MTPs, allocating portions of the bandwidth not utilized by the considered MTP. Therefore, $C_M$ should be interpreted as a parameter that provides an estimate of the overall performance potential of the MTP.}}
	{Considering the system capacity in \eqref{eq:objFunct}, the objective function for the optimization problem }corresponds to the problem of maximizing the total capacity over the considered angle/frequency interval.
	\subsubsection{Linear Constraint}
	In the following, we consider an \ac{MTP} based on the mapping in \eqref{eq:Meta_Davide}, which allows for nearly ideal implementation through Foster circuits. As in \eqref{eq:Meta_Davide}, the following linear constraint is imposed in the optimization problem 
	\begin{equation}
		\label{eq:linConstr}
		\psi_{k,n} = \an(\fk-f_0)+ \gn,
	\end{equation}
	where $\alpha_n$ and $\gamma_n$ are calculated for each element $n$ of the \ac{MTP} and thus, become the optimization variables of the problem. Note that, in the ideal \ac{MTP} model case \eqref{eq_model_1}, the optimization process is expected boil down to the solution in {\eqref{eq:reflectionCoeffOurModel}, i.e., $\alpha_n = \frac{2\pi\nu_n\Delta_v}{\lambda} \alpha$ and $\gamma_n = \frac{2\pi\nu_n\Delta_v}{\lambda}\gamma$}. 
	\subsubsection{Optimization Problem}
	From \eqref{eq:objFunct} and \eqref{eq:linConstr}{, indicating $\boldsymbol{\alpha} = [\alpha_1, \ldots, \alpha_N]$ and $\boldsymbol{\gamma} = [\gamma_1, \ldots, \gamma_N]$,} the optimization problem at hand can be formulated as
	\begin{align}
		\label{eq:GammaOptProb}
		&\max_{\boldsymbol{\alpha}, \boldsymbol{\gamma}} {C}_M\\
		& \text{s.t}~\psi_{k,n} = \an(f_k-f_0)+ \gn. \nonumber
	\end{align}
	
	Problem \eqref{eq:GammaOptProb} is very complex, mainly due to the non-linear (and non-convex) nature of the objective function, which involves matrix inversions, as shown in the channel model in \eqref{model_RIS_MP}. To address this issue, an iterative approach based on AO is employed.
	
	To elaborate, {for simplicity of notation, let introduce $\boldsymbol{\Delta}_k \in \mathbb{C}^{N \times N} = \boldsymbol{\Gamma}_k^{-1} = \left\{e^{-\jmath \psi_{k,n}}\right\}$ and consider $\boldsymbol{\Delta}_{k,-n}$ as the matrix $\boldsymbol{\Delta}_k$ with the $n$-th element of the diagonal set to zero, i.e.:}  
	\begin{equation}
		\label{eq:GammaPrimo}
		\{\boldsymbol{\Delta}_{k,-n}\}_{l,l} = 
		\begin{cases}
			\{\boldsymbol{\Delta}_k\}_{l,l} & if ~ l \neq n\\
			0 & if ~ l = n
		\end{cases}.
	\end{equation}
	{It is therefore possible to write the matrix $\boldsymbol{\Delta}_k$ as the sum of a diagonal matrix and a rank-one matrix, namely:}
	\begin{equation}
		\label{eq:Sh1}
		\boldsymbol{\Delta}_k =\boldsymbol{\Delta}_{k,-n}+e^{-\jmath\psi_{k,n}}\mathbf{e}_{n}\mathbf{e}_{n}^T,
	\end{equation}
	for $n=1,\ldots,N $, where $\mathbf{e}_{n}$ is an all-zero vector except in position $n$ where it is one. Hence, denoting by
	\begin{align}
		\label{eq:A_B_c}
		&\mathbf{A}_{k,-n} =  \left[\boldsymbol{\Delta}_{k,-n} - \mathbf{S}_{SS}^{(k)}\right]^{-1} \nonumber \\
		&\mathbf{B}_{k,-n} =  \mathbf{A}_{k,-n} \mathbf{e}_{n} \mathbf{e}^T_{n}  \mathbf{A}_{k,-n}\\
		&c_{k,-n} =  \mathbf{e}^T_{n} \mathbf{A}_{k,-n} \mathbf{e}_{n}, \nonumber
	\end{align} from the Sherman Morrison inversion formula \cite{Sherman1950Adjustment} we can write for the $n$-th element:
	\begin{equation}\label{eq:Sh2}
		\left(\boldsymbol{\Gamma}_k^{-1}-\mathbf{S}_{SS}^{(k)}\right)^{-1} = \mathbf{A}_{k,-n} - \frac{\mathbf{B}_{k,-n}}{{e^{\jmath\psi_{k,n}}}+c_{k,-n}}.
	\end{equation}
	{The expression} in \eqref{eq:Sh2} allows the problem \eqref{eq:GammaOptProb} to be divided into subproblems, one for each element $n$ of the \ac{MTP}. For this purpose, it is convenient to introduce the vector $\boldsymbol{\psi}_{n} \in \mathbb{R}^{K \times 1}$ defined as {the vector containing all the $K$ phases of the reflection coefficient of the $n$-th element of the \ac{MTP} for all the frequencies $f_k$, i.e., }$\boldsymbol{\psi}_{n} = [{\psi}_{1,n}, \ldots, {\psi}_{K,n}]^T$. Then, letting {$\boldsymbol{\Psi} \in \mathbb{R}^{K \times N}$} be the matrix obtained by stacking the vectors $\boldsymbol{\psi}_{n}$ in separate columns, i.e., { $\boldsymbol{\Psi}$}$ = [\boldsymbol{\psi}_{1}, \ldots, \boldsymbol{\psi}_{N}]$, it is easy to see that the terms in \eqref{eq:A_B_c}, for any value of $k$, depend on the matrix{ $\boldsymbol{\Psi}$} except its $n$-th column $\boldsymbol{\psi}_{n}$. Therefore, for convenience, {we denote { $\boldsymbol{\Psi}_{-n} $$ \in \mathbb{R}^{K \times (N - 1)}$} as the matrix obtained by removing the $n$-th column $\boldsymbol{\psi}_{n}$ from $\boldsymbol{\Psi}$.}
	The $n$-th subproblem of the original problem corresponds to maximizing \eqref{eq:objFunct} with respect to $\boldsymbol{\psi}_{n}$, given a certain{ $\boldsymbol{\Psi}_{-n}$}. By virtue of the linear constraint \eqref{eq:linConstr}, on the other hand, it can be seen that $\boldsymbol{\psi}_{n}$ depends only on the parameters $\alpha_n$ and $\gamma_n$. Therefore, using \eqref{eq:Sh2}, we can rewrite \eqref{model_RIS_MP} as function of the $n$-th element only, i.e., 
	\begin{align}
		\tilde{h}^{(k)}(\Psi_{k,n}) = & 
		~ \mathbf{s}_{RM}^{(k)}\left(\mathbf{A}_{k,-n} - \frac{\mathbf{B}_{k,-n}}{{e^{\jmath\Psi_{k,n}}}+c_{k,-n}}\right)\mathbf{s}_{MT}^{(k)} + \mathbf{s}_{RT}^{(k)} 
	\end{align}
	
	{For convenience of notation, it is useful at this stage to introduce the following parameters:}
	\begin{align}
		\label{eq:Parameters}
		& D_{n}^{(k)} = \mathbf{s}_{RT}^{(k)} + \mathbf{s}_{RM}^{(k)} \mathbf{A}_{k,-n} \mathbf{s}_{MT}^{(k)} \\
		& F_{n}^{(k)} = -\mathbf{s}_{RM}^{(k)} \mathbf{B}_{k,-n} \mathbf{s}_{MT}^{(k)}, \nonumber
	\end{align} 
	so that \eqref{model_RIS_MP} can be rewritten as
	\begin{align}
		\tilde{h}^{(k)}(\Psi_{k,n}) =   D_{n}^{(k)}+\frac{F_{n}^{(k)}}{{e^{\jmath\left(\an(f_k-f_0)+ \gn\right)}}+c_{k,-n}}
	\end{align}
	and the $n$-th subproblem of \eqref{eq:GammaOptProb} as
	\begin{align}
		\label{eq:GammaOptProb3}
		&\max_{\alpha_n, \gamma_n} C_M(\alpha_n, \gamma_n) =  \sum_{k = 1}^ K \frac{W}{K} \text{log}\left(1+ \frac{\lvert\tilde{h}^{(k)}(\Psi_{k,n})\rvert^2 P_t}{ W 
			N_0}\right) \nonumber\\
		& \text{s.t}~\Psi_{k,n} = \an(f_k-f_0)+ \gn ~ \text{for} ~ k = 1,\ldots,K.
	\end{align}
	Although the problem remains non-convex, it has been simplified to maximizing a function of two real variables, $\alpha_n$ and $\gamma_n$. This allows for a solution through exhaustive search by discretely sampling the domain of $\alpha_n$ and $\gamma_n$. The optimal value for $\gamma_n$ can be searched within a set $\mathcal{I}_{\gamma}$ of $N_{\gamma}$ points obtained through uniform sampling of the interval $\{0, 2\pi\}$. As for $\alpha_n$, the solution belongs to the discrete set $\mathcal{I}_{\alpha}$ of $N_{\alpha}$ points, sampled in the domain $ \{\tilde{\alpha}^{(\text{min})}, \tilde{\alpha}^{(\text{max})}\}$. The extremes of the interval represent the minimum and maximum slope of the ideal case, i.e., $\tilde{\alpha}^{(\text{min})} = \frac{2\pi\nu_1 \Delta_{\nu}}{\lambda_0}$ and  $\tilde{\alpha}^{(\text{max})} = \frac{2\pi\nu_I \Delta_{\nu}}{\lambda_0}$. 
	
	{The algorithm proceeds iteratively, with an outer loop indexed by $q$, in which the matrix{ $\boldsymbol{\Psi}^{(q)}$} is computed. For each $q$, there is an inner loop indexed by $n$, in which{ $\boldsymbol{\Psi}_{-n}^{(q)}$} is used to calculate the terms in \eqref{eq:A_B_c} and, consequently, the parameters in \eqref{eq:Parameters} for each $k = 1, \ldots, K$. Then, the problem \eqref{eq:GammaOptProb3} is solved,{ $\boldsymbol{\psi}_n$} is computed from \eqref{eq:linConstr}, and the new value of{ $\boldsymbol{\psi}_n$} is substituted back into{ $\boldsymbol{\Psi}^{(q)}$} to proceed to the next $n$.} The entire procedure is detailed in Algorithm \ref{alg:optimalGamma}. 
	
	{Note that in each iteration, it is found a value of $\boldsymbol{\Psi}_n$ that either increases the objective function or keeps it unchanged, ensuring that the proposed algorithm converges to a local optimum.} {As for the initial condition, we start setting {$\alpha_n = \frac{2\pi\nu_n\Delta_v}{\lambda_0} \alpha$ and $\gamma_n = \frac{2\pi\nu_n\Delta_v}{\lambda_0}\gamma$}} where $\alpha$ and $\gamma$ are those of the ideal \ac{MTP} case given in \eqref{eq:Meta_Davide2}. In this way, given the nature of the algorithm to always produce a non-decreasing solution of the objective function, in the case of the ideal \ac{MTP} model \eqref{eq_model_1} the algorithm will inevitably remain at the initial value, i.e., it will exit with the optimal solution. 
	
	\begin{algorithm}
		\small
		\SetAlgoLined
		Set $q = 1$, $\boldsymbol{\Psi}^{(1)}= [\Psi_1,\ldots, \Psi_N]$;\\ 
		Compute $C_M^{(1)}$ from $\boldsymbol{\Psi}^{(1)}$ \eqref{eq:objFunct} and $\{\tilde{\alpha}_n^{(\text{min})}, \tilde{\alpha}_n^{(\text{max})}\}$;\\
		Set $\mu =1$ and an arbitrarily small value $\varepsilon << \mu$, ;\\ 
		\For{$\tk \in \mathcal{T}_K$}{
			Compute $\mathbf{s}_{RT}^{(k)}$, $\mathbf{s}_{RM}^{(k)}$, $\mathbf{S}_{SS}^{(k)}$, $\mathbf{s}_{MT}^{(k)}$ according to \cite{abrardo2023design};
		}
		\While{$\mu > \varepsilon$}{
			\For{$n = 1,\ldots, N$}{
				Store $\boldsymbol{\Psi}_{-n}^{(q)}$;\\
				\For{$\tk \in \mathcal{T}_k$}{
					Compute $\boldsymbol{\Delta}_{k,-n}$ \eqref{eq:GammaPrimo}, $\mathbf{A}_{k,-n}$, $\mathbf{B}_{k,-n}$, $c_{k,-n}$ \eqref{eq:A_B_c}, $D_n^{(k)}$, $F_n^{(k)}$, \eqref{eq:Parameters}
				}
				\For{$\an \in \mathcal{I}_\alpha$}{
					\For{$\gn \in \mathcal{I}_\gamma$}{
						Compute ${C}_M(\an, \gn)$;\\
					}
				}
				Set $(\an^\star, \gn^\star) = \arg\max\limits_{\alpha_n, \gamma_n}{C}_M(\an, \gn)$ and ${C}_M^{(q+1)} = {C}_M(\an^\star, \gn^\star)$;\\
				Update $\boldsymbol{\psi}_{n}$ and $\boldsymbol{\Psi}^{(q)}$;\\
			}
			$\mu = \frac{\lvert {C}_M^{(q+1)}-{C}_M^{(q)}\rvert}{\lvert{C}_M^{(q)}\rvert}$, 
			$q = q +1$ ; \\
		}
		\caption{Reflection coefficient optimization for realistic \ac{MTP} models }
		\label{alg:optimalGamma}
	\end{algorithm}
	
	\begin{figure}
		\vspace{-\baselineskip}
		\centering
		\includegraphics[width = 0.7\linewidth]{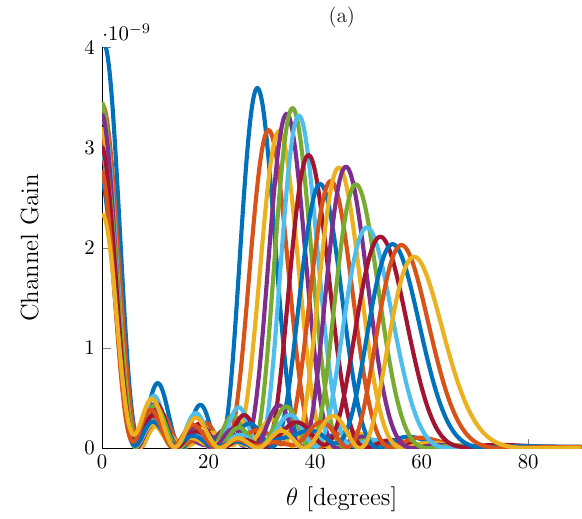}\\
		\includegraphics[width = 0.7\linewidth]{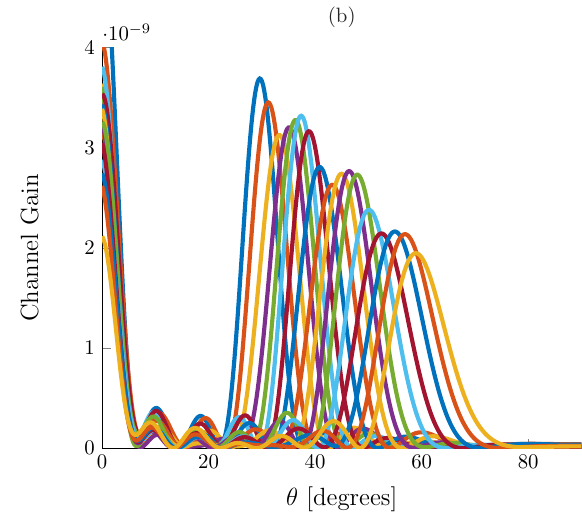}
		
		\caption{Channel gain with the Foster's first form circuit design and realistic \ac{MTP} model with optimized reflection coefficient, for $\Delta_y = \lambda_0/2$ (a) and $\Delta_y = \lambda_0/4$ (b).  }
		\label{fig:ChGainGammaOpt}
		\vspace{-\baselineskip}
	\end{figure}
	\subsubsection{Computational Complexity}
	{The computational complexity of the proposed algorithm for each iteration $q$ consists of two parts. The first is the solution of problem \eqref{eq:GammaOptProb3} for $n = 1,\ldots,N$, which requires for each $n$ the evaluation of the capacity in \eqref{eq:GammaOptProb3} $N_{\gamma} N_{\alpha}$ times, with a complexity of the order of $\mathcal{O}(NK N_{\gamma} N_{\alpha})$. The second part involves calculating the terms in \eqref{eq:A_B_c} for each $k$ and for each $n$. In the calculation of these terms, we assume for simplicity that the product of an $n \times p$ matrix by a $p \times q$ matrix entails a number of operations proportional to $npq$, and we adopt the customary assumption that the computational cost of computing the inverse of a matrix depends on the cube of the number of its elements. Accordingly, the complexity is $\mathcal{O}(2N^4 K)$. If we denote by $N_t$ the number of iterations on $q$ to reach convergence, the complexity then becomes $\mathcal{O}(N_t NK N_{\gn} N_{\alpha_n}) + \mathcal{O}(2N_tN^4 K)$. This value is in line with the complexities generally encountered when dealing with RIS, which are always of the order of $N^r$ with $r > 3$. It should be noted, however, that since an \ac{MTP} can operate in static or quasi-static mode, the reconfiguration requirements can be relaxed compared to those of an RIS, assuming that the considered system setting is static and does not require any reconfigurability.

		\section{Results and Comparisons}
		\label{sec:Results}
		To illustrate the effectiveness of the optimization procedure shown in Algorithm \ref{alg:optimalGamma}, we replicate the same setup as that shown in Figures \ref{fig:ChGainFosterReactanceRealRISNoSpecular}a and \ref{fig:ChGainFosterReactanceRealRISNoSpecular}b. The coefficients $\Gamma_n(f_k)$ are obtained through the Foster circuit as detailed in Algorithm \ref{alg:optimalLC} while the phases are obtained through Algorithm \ref{alg:optimalGamma}. For the simulations we consider $P_{t} = 0~ \text{dBm}$ and $N_0 = -165.37~ \text{dBm}$. The results are shown in Figures \ref{fig:ChGainGammaOpt}a and \ref{fig:ChGainGammaOpt}b for $\Delta_y = \lambda_0/2$ and $\Delta_y = \lambda_0/4$, respectively. In particular, for the implementation of Algorithm \ref{alg:optimalGamma}, we consider $N_{\alpha} = 300$ and $N_{\gamma} = 100$. The results show that the proposed optimization improves the performance of the MTP, an effect that is particularly evident in the case $\Delta_y = \lambda_0/4$, as expected.\begin{figure}
			\vspace{-\baselineskip}
			\centering
			\includegraphics[width=0.9\linewidth]{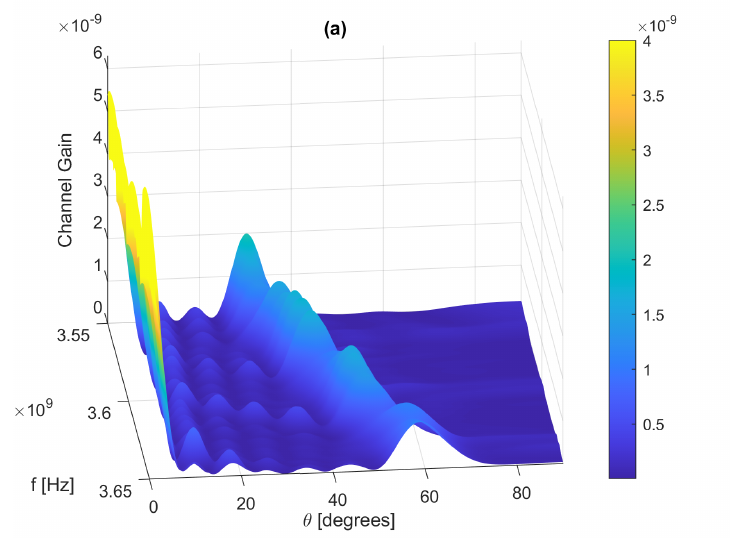}\\
			\includegraphics[width=0.9\linewidth]{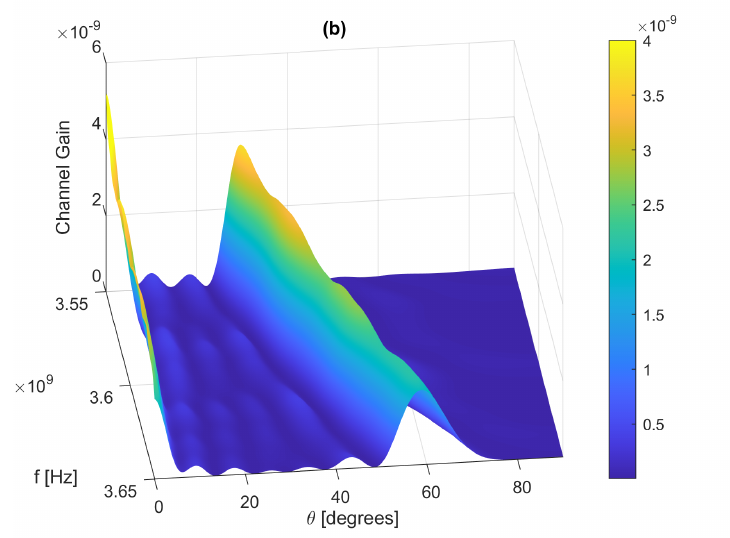}
			\caption{Channel gain comparison for $\Delta_y = \lambda_0/4$ in case of non optimized (a) and optimized (b) reflection coefficient for $\theta \in \mathcal{T}$ and $f \in \mathcal{F}$. }
			\label{fig:3Dgraph}
			\vspace{-\baselineskip}
		\end{figure}
		Figures \ref{fig:3Dgraph}a and \ref{fig:3Dgraph}b show the transfer function $|h(\theta,f)|^2$ for $\Delta y = \lambda_0/4$ in the two cases where the reflection coefficients are not optimized and are optimized, respectively. 
		In this case, the performance improvement due to optimization carried out according to Algorithm \ref{alg:optimalGamma} is evident.\\
		{To confirm the results obtained by applying the multiport network model \cite{DR1}, a full-wave simulator is also used to simulate the scenario in Figure \ref{fig:ChGainGammaOpt}a} with the corresponding parameters reported in Table \ref{tab:parameters}. {To this end, we consider an MTP modeled as a uniform planar array of nearly-resonant loaded dipoles.} 
		\begin{figure}
			
			\centering
			\includegraphics[width=0.7\linewidth]{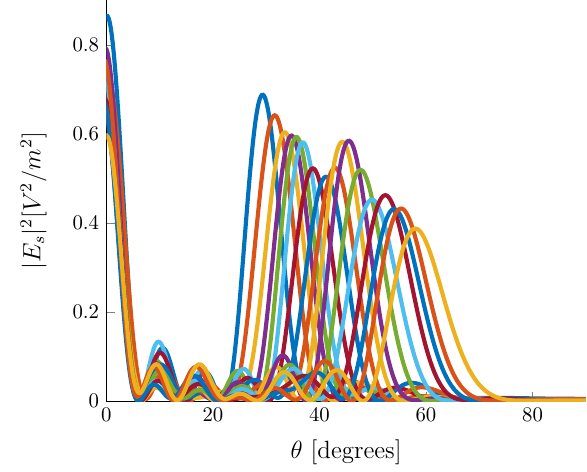}
			\caption{Scattered squared electric field strength with the Foster's first form circuit design and realistic \ac{MTP} model with optimized reflection coefficient, for $\Delta_y = \lambda_0/2$, and full-wave simulator}
			\label{fig:Feko}
			\vspace{-\baselineskip}
		\end{figure}
		At each configuration corresponding to a given receiver position, the scattering matrix is computed and used in the optimization process to optimize the unit cells of the MTP (see Section \ref{sec:ReflectionCoefficientOptimization}).
		The SPICE circuits of the static loads that determine the unit cell structure are obtained according to the procedure described in Section \ref{sec:CircuitModelDesign}.
		These circuits models are then imported in the EM simulator. The complete unit cells of the MTP are obtained by connecting the Foster circuits to the ports of thin dipoles, which are the considered scatterers of the MTP. 
		Finally, the full-wave numerical simulation is carried out with a plane wave excitation and the scattered electric field is computed at each operating frequency. In Figure \ref{fig:Feko} the squared amplitude of the scattered electric field versus $\theta$, is reported. It is found that the obtained curves exhibit the same behavior of those in Figure \ref{fig:ChGainGammaOpt}a, which confirm the performance of the proposed optimization procedure. 
		
		In addition, we analyze $C_M$ defined in \eqref{eq:objFunct} and normalized with respect to the bandwidth $W$. The gains $h_M(k)$ are computed using the realistic model \eqref{eq_model_2} while the reflection coefficients $\Gamma_n(f_k)$ are obtained through the Foster circuit obtained according to Algorithm \ref{alg:optimalLC}. This case is referred to as \textit{MTP} for the case where the phases of the reflection coefficients are obtained through Algorithm \ref{alg:optimalGamma} or \textit{Non-Opt MTP} for the case where the phases are calculated as in \eqref{eq:reflectionCoeffOurModel}. As a baseline solution, we consider the case where the MTP optimization is carried out using Algorithm \ref{alg:optimalGamma} without imposing the phase linearity constraint in \eqref{eq:Sh2}, i.e., assuming $\psi_{k,n} \in [0,2\pi]$. This corresponds to having an ideal load network that can achieve any phase shift at any frequency, allowing each beam related to the angle $\theta_k$ to be optimized independently of the others.  
		This case is defined as {\textit{NC MTP } (non constrained)}. The results are reported for two angular sets $\mathcal{T}_A = [\pi/4, \pi/2]$ and $\mathcal{T}_B = [\pi/6, \pi/3]$.
		The bar chart in Figure \ref{fig:Capacity} highlights the gain due to optimization. Additionally, it is evident that, unlike the ideal case, the use of an MTP with the realistic \ac{MTP} model incurs a small performance penalty compared to an {non-constrained MTP}, due to the phase linearity constraint that must be imposed for its feasibility. 
		{However, this performance degradation is a compromise for the device realizability. In Figure \ref{fig:phShiftComparison} we report the comparison, for a given element $n$ of the optimized phase shift obtained for the \textit{NC-MTP} and \textit{MTP} cases for $\theta_k \in \mathcal{T}_B$ and $\Delta_y = \lambda_0/2$. It is evident that the non-constrained case actually entails a phase behavior that cannot be {easily} obtained by means of the considered {synthesis procedure for the reactance}, as proposed in Section \ref{sec:CircuitModelDesign} for the \textit{MTP} case. } 
		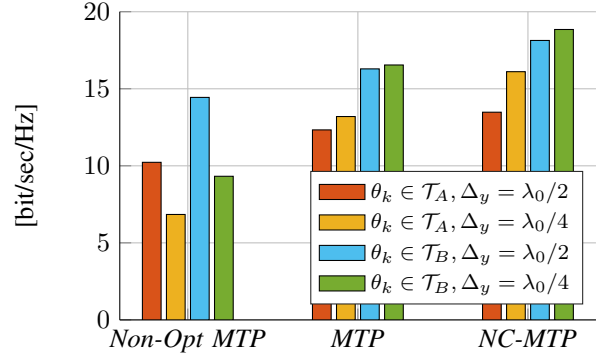
\begin{figure}
			\vspace{-\baselineskip}
			\centering
%
%
\definecolor{mycolor1}{rgb}{0.00000,0.44700,0.74100}%
\definecolor{mycolor2}{rgb}{0.85000,0.32500,0.09800}%
\definecolor{mycolor3}{rgb}{0.92900,0.69400,0.12500}%
\definecolor{mycolor4}{rgb}{0.49400,0.18400,0.55600}%

\definecolor{mycolor1}{rgb}{0.85000,0.32500,0.09800}%
\definecolor{mycolor2}{rgb}{0.92900,0.69400,0.12500}%
\definecolor{mycolor3}{rgb}{0.49400,0.18400,0.55600}%
\definecolor{mycolor4}{rgb}{0.46600,0.67400,0.18800}%
\definecolor{mycolor3}{rgb}{0.30100,0.74500,0.93300}%
\definecolor{mycolor7}{rgb}{0.63500,0.07800,0.18400}%
\begin{tikzpicture}

\begin{axis}[%
width = 0.9\linewidth,
height=0.65\linewidth,
bar shift auto,
xmajorgrids,
ymajorgrids,
xmin=0.6,
xmax=3.4,
xtick=data,
xticklabels={\textit{Non-Opt MTP}, \textit{MTP}, \textit{NC-MTP}},
ymin=0,
ymax=20,
ylabel= {[bit/sec/Hz]},
axis background/.style={fill=white},
axis x line*=bottom,
axis y line*=left,
legend style={at={(0.4,0.05)}, anchor=south west, legend cell align=left, align=left, draw=white!15!black, font = \footnotesize}
]
\addplot[ybar, bar width=7, fill=mycolor1, draw=black, area legend] table[row sep=crcr] {%
1	10.2275383899169\\
2	12.3315487471077\\
3	13.4800697584016\\
};
\addlegendentry{$\theta_k \in \mathcal{T}_A, \Delta_y = \lambda_0/2$}

\addplot[ybar,  bar width=7, fill=mycolor2, draw=black, area legend] table[row sep=crcr] {%
1	6.84286794998356\\
2	13.1991123112313\\
3	16.1097812794413\\
};
\addlegendentry{$\theta_k \in \mathcal{T}_A, \Delta_y = \lambda_0/4$}

\addplot[ybar,  bar width=7, fill=mycolor3, draw=black, area legend] table[row sep=crcr] {%
1	14.4405292550937\\
2	16.2915143348642\\
3	18.1368140681772\\
};
\addlegendentry{$\theta_k \in \mathcal{T}_B, \Delta_y = \lambda_0/2$}

\addplot[ybar,  bar width=7, fill=mycolor4, draw=black, area legend] table[row sep=crcr] {%
1	9.3217745307927\\
2	16.5447416248683\\
3	18.8486276242058\\
};
\addlegendentry{$\theta_k \in \mathcal{T}_B, \Delta_y = \lambda_0/4$}

\end{axis}

\end{tikzpicture}%
			\caption{Bandwidth-normalized system capacity analysis for the MTP in case of non optimization, optimization with linear constraints and optimization without constraints,  with $\mathcal{T}_A = [\pi/4,\pi/2]$ and $\mathcal{T}_B = [\pi/6,\pi/3]$}
			\label{fig:Capacity}
			\vspace{-\baselineskip}
		\end{figure}
		
		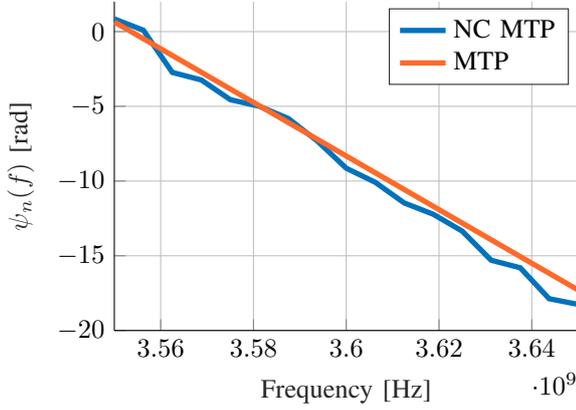
\begin{figure}
			\centering
%
%
\definecolor{mycolor1}{rgb}{0.00000,0.44706,0.74118}%
\definecolor{mycolor2}{rgb}{1.00000,0.41176,0.16078}%
\begin{tikzpicture}

\begin{axis}[%
width=0.7\linewidth,
height=0.5\linewidth,
at={(0.756in,0.48in)},
scale only axis,
xmin=3550000000,
xmax=3650000000,
ymin=-20,
ymax=2,
axis background/.style={fill=white},
axis x line*=bottom,
axis y line*=left,
xmajorgrids,
ymajorgrids,
ylabel style={font=\color{white!15!black}},
ylabel = {$\psi_n(f)$ [rad]}, 
xlabel style={font=\color{white!15!black}},
xlabel = {Frequency [Hz]}, 
legend style={legend cell align=left, align=left, draw=white!15!black}
]
\addplot [color=mycolor1, line width=2.0pt]
  table[row sep=crcr]{%
3550000000	0.848227984043134\\
3556250000	0.0942477762664457\\
3562500000	-2.73317951547969\\
3568750000	-3.23584193561313\\
3575000000	-4.55531848529313\\
3581250000	-4.99513785824116\\
3587500000	-5.81194680499683\\
3593750000	-7.31990750216148\\
3600000000	-9.14203024664438\\
3606250000	-10.0845211957116\\
3612500000	-11.4668170775792\\
3618750000	-12.2207955826716\\
3625000000	-13.3517671209051\\
3631250000	-15.2995501298389\\
3637500000	-15.8022125499723\\
3643750000	-17.8756650964589\\
3650000000	-18.2526540949169\\
};
\addlegendentry{NC MTP}

\addplot [color=mycolor2, line width=2.0pt]
  table[row sep=crcr]{%
3550000000	0.643250555867715\\
3556250000	-0.47780707903564\\
3562500000	-1.59885827558807\\
3568750000	-2.71991924093449\\
3575000000	-3.84099359967902\\
3581250000	-4.96204935114509\\
3587500000	-6.08310255202295\\
3593750000	-7.20415901046887\\
3600000000	-8.32521016783805\\
3606250000	-9.4462798818326\\
3612500000	-10.567348843513\\
3618750000	-11.6883997735063\\
3625000000	-12.8094564988209\\
3631250000	-13.9305093380538\\
3637500000	-15.0515658173755\\
3643750000	-16.1726403630182\\
3650000000	-17.2937004585154\\
};
\addlegendentry{MTP}

\end{axis}
\end{tikzpicture}%
			\caption{{Phase shift comparison for the \textit{NC-MTP} and \textit{MTP} cases, for $\theta_k \in \mathcal{T}_B$ and $\Delta_y = \lambda_0/2$.}}
			\label{fig:phShiftComparison}
		\end{figure}

		\section{Conclusions}
		\label{sec:Conclusions}
		This work presented a comprehensive examination of \ac{MTP} technology, emphasizing its potential as a passive solution for enhancing wireless communications through frequency-dependent beamsteering and focusing. By refining the design criteria and employing an accurate multiport network model, we addressed the limitations of previous studies that relied on oversimplified linear phase models, thereby improving the feasibility and performance of MTPs. Our proposed optimization techniques for circuit implementation and non-idealities are demonstrated to be effective through the use of a full-wave simulator. 
		
		\bibliographystyle{IEEEtran}
		\bibliography{references}
	\end{document}